%% file: main.tex
\pdfoutput=1

\documentclass{article}

\usepackage[preprint]{neurips_2026}

\usepackage[T1]{fontenc}

\usepackage[utf8]{inputenc}

\usepackage{microtype}

\usepackage{paralist}
\usepackage{xspace}
\usepackage{makecell}
\usepackage{booktabs}
\usepackage{amsmath}
\usepackage{amssymb} 
\usepackage{pifont} 
\usepackage{dblfloatfix}
\usepackage{enumitem}
\usepackage{multirow}
\usepackage{rotating}
\usepackage{array}
\usepackage[autostyle=true]{csquotes}
\usepackage{inconsolata}

\usepackage{colortbl}
\usepackage{xcolor}
\definecolor{Red}{RGB}{192, 0, 0}
\definecolor{Blue}{RGB}{13, 110, 158}
\definecolor{Yellow}{RGB}{218, 169, 20}
\definecolor{lightyellow}{RGB}{255,255,153}

\definecolor{HighlightBlue}{RGB}{0, 100, 148}
\definecolor{HighlightRed}{RGB}{230, 57, 70}

\definecolor{LightRed}{HTML}{ffe0e0}
\definecolor{LightBlue}{HTML}{def5ff}
\definecolor{LightYellow}{HTML}{FFF6DB}
\definecolor{LightGreen}{HTML}{eff9f0}

\usepackage{hyperref}
\usepackage{url}
\usepackage{pifont}
\usepackage{soul}
\usepackage{color, xcolor}

\RequirePackage[most]{tcolorbox}
\RequirePackage{titletoc}

\usepackage{listings}

\definecolor{listingbg}{RGB}{246,243,238}      
\definecolor{listingframe}{RGB}{205,195,182}   
\definecolor{pyorange}{RGB}{200,85,25}         
\definecolor{pygreen}{RGB}{35,120,60}          
\definecolor{pystring}{RGB}{30,110,55}         

\lstdefinestyle{pythonstyle}{%
    language=Python,
    basicstyle=\small\ttfamily\color{black!90},
    keywordstyle=\color{deepblue}\bfseries,
    stringstyle=\color{pystring},
    commentstyle=\color{deepgray}\itshape,
    identifierstyle=\color{black!85},
    backgroundcolor=\color{listingbg},
    showstringspaces=false,
    breaklines=true,
    breakatwhitespace=false,
    columns=flexible,
    keepspaces=true,
    aboveskip=6pt, belowskip=6pt,
    frame=single,
    rulecolor=\color{listingframe},
    framerule=0.6pt,
    xleftmargin=10pt,
    xrightmargin=6pt,
    framexleftmargin=6pt,
    framexrightmargin=4pt,
    tabsize=4,
    numberstyle=\normalfont\tiny\color{deepgray},
    numbersep=6pt,
}
\lstdefinestyle{promptstyle}{%
    basicstyle=\scriptsize\ttfamily,
    breaklines=true,
    breakatwhitespace=false,
    columns=flexible,
    keepspaces=true,
    aboveskip=2pt, belowskip=2pt,
}

\usepackage{wrapfig,lipsum,booktabs}

\usepackage{colortbl}
\definecolor{lightyellow}{RGB}{255,242,204}
\definecolor{lightorange}{RGB}{251,229,214}
\definecolor{lightgreen}{RGB}{226,240,217}
\definecolor{lightblue}{RGB}{222,235,247}
\definecolor{lightgray}{RGB}{209,201,206}
\definecolor{deepgray}{RGB}{178,164,173}
\definecolor{deepblue}{RGB}{70,130,180}

\usepackage{threeparttable}
\usepackage{xspace}
\usepackage{graphicx}
\usepackage{adjustbox}
\usepackage{float}
\usepackage{booktabs}
\usepackage{multicol}
\usepackage{multirow}
\usepackage[ruled,boxed,linesnumbered]{algorithm2e}
\usepackage{caption}
\usepackage{cleveref}
\usepackage{subcaption}
\crefname{figure}{Fig.}{Figs.}
\Crefname{figure}{Fig.}{Figs.}
\crefname{table}{Tab.}{Tabs.}
\Crefname{table}{Tab.}{Tabs.}
\crefname{section}{Sec.}{Secs.}
\Crefname{section}{Sec.}{Secs.}
\crefname{appendix}{Appendix}{Appendix}
\Crefname{appendix}{Appendix}{Appendix}

\definecolor{darkgreen}{rgb}{0.0, 0.5, 0.0} 
\definecolor{darkred}{rgb}{0.6, 0.0, 0.0}   

\newcommand{\checkyes}{\textcolor{white}{\setlength{\fboxsep}{1pt}\colorbox{darkgreen}{\ding{51}}}} 
\newcommand{\crossno}{\textcolor{white}{\setlength{\fboxsep}{1pt}\colorbox{darkred}{\ding{55}}}} 

\newcommand{\tool}{RustPrint\xspace}
\newcommand{\code}[1]{{\small\texttt{#1}}}
\NewDocumentCommand{\revanth}
{ mO{} }{\textcolor{blue}{\textsuperscript{\textit{Revanth}}\textsf{\textbf{\small[#1]}}}}
\usepackage{booktabs,multirow,graphicx,colortbl,xcolor}

\definecolor{rowgray}{gray}{0.88}
\definecolor{rowgraylight}{gray}{0.93}
\definecolor{rowgraydark}{gray}{0.72}

\title{Documentation-Guided Agentic Codebase Migration from C to Rust}

\author{%
  Minh Le-Anh \\
  FPT Software AI Center \\
  Ha Noi, Viet Nam \\
  \texttt{minhla4@fpt.com}
  \And
  Anh Nguyen Hoang \\
  FPT Software AI Center \\
  Ha Noi, Viet Nam \\
  \texttt{anhnh2220@fpt.com}
  \AND
  Bach Le \\
  University of Melbourne \\
  Melbourne, Australia \\
  \texttt{bach.le@unimelb.edu.au}
  \And
  Nghi D. Q. Bui \\
  Center of AI Research, VinUniversity \\
  Ha Noi, Viet Nam \\
  \texttt{bdqnghi@gmail.com}
}

\begin{document}
\maketitle

\input{sections/abstract.tex}
\input{sections/introduction}
\input{sections/related_work}
\input{sections/migraagent}
\input{sections/evaluation_design}
\input{sections/experiment}
\section*{Discussion \& Limitations}
\label{sec:limitation}
\input{sections/limitations}
\input{sections/conclusions}

\newpage
\bibliographystyle{plainnat}
\bibliography{main}


\newpage

\newpage
\appendix
\label{sec:Appendix}
\input{sections/appendix}

\newpage
\input{checklist}

\end{document}

%% file: sections/abstract.tex
\begin{abstract}

Migrating legacy C repositories to Rust promises stronger memory safety, but existing translators often work at the level of files or functions and miss architectural intent. We present \tool, a documentation-guided agentic framework for repository-level C-to-Rust migration. \tool first converts the source repository into architecture-aware documentation and treats it as a migration blueprint capturing module structure, data flow, APIs, and design rationale. Coding agents then use this blueprint to plan crates, implement modules, check compilability, reduce unsafe code, and iteratively refine the translated repository. \tool next compares documentation from the Rust output against the source documentation and uses mismatches as repair signals. It also translates and runs source test suites so runtime failures can guide targeted fixes. Experiments on eight real-world C repositories ranging from 11K to 84K LoC show that \tool\ compiles every target under both an open-weight (Kimi-K2-Instruct) and a closed-weight (GPT-5.4) backbone, while prior LLM-based translators (Self-Repair, EvoC2Rust) fail repository-wide. With the open-weight Kimi-K2-Instruct backbone, \tool\ exceeds an agentic Claude Code baseline on feature preservation (\textbf{93.26\%} vs.\ \textbf{52.52\%}) and on cross-evaluation test pass rate (\textbf{95.17\%} vs.\ \textbf{79.85\%}). These results suggest that documentation-guided coordination is a useful direction for scalable codebase migration.

\end{abstract}

%% file: sections/introduction.tex
\section{Introduction}
\label{sec:introduction}

The C programming language remains the foundation of systems software, including kernels, embedded firmware, networking stacks, and safety-critical infrastructure \cite{Peta2022CPL}. This legacy carries a persistent cost: memory safety vulnerabilities rooted in C's pointer model continue to drive critical exploits \cite{Sze2013SoK, van2023MeMS, Seidel2024BringingRTAC}. Rust offers memory safety and data-race freedom without giving up low-level control \cite{jung2021Sspr, Panter_2024}, and has gained adoption in systems settings such as Linux, Android, and embedded software \cite{Seidel2024BringingRTAC, li2024rfl, may2021apsm}. The central obstacle is scale. Rewriting mature C repositories by hand is expensive, and controlled studies show that even experienced developers find C-to-Rust migration cognitively demanding \cite{Li2024TranslatingCTJ}.

Recent progress in LLMs for coding has made repository-scale software engineering more plausible. Multi-agent coding systems can now plan, edit, test, and iteratively repair code across long-horizon tasks \cite{hong2024metagpt, qian2024chatdev, nguyen2024agilecoder, islam2024mapcoder}. This trend suggests that C-to-Rust migration should be treated not as isolated translation, but as a coding-agent task over an existing repository. However, current migration systems still operate mostly at the level of functions or files, or coordinate translation through dependency graphs, build order, and generated skeletons. These strategies help with compilation, but they do not explicitly transfer the architecture, intent, and behavioral contracts that shape the repository as a whole.

Real-world migration is fundamentally a codebase-understanding problem. A repository contains shared data structures, module boundaries, naming conventions, build assumptions, cross-file invariants, and design rationale that cannot be recovered from isolated functions alone. Human developers first build a mental model of the system before rewriting it in another language. This observation motivates a different migration interface: use documentation as a whole-codebase intermediate representation that captures the source repository before agents generate and revise the target implementation.


In this paper, we present \textbf{\tool}, a framework for repository-level C-to-Rust migration that introduces a documentation-driven paradigm for idiomatic code translation. Rather than translating code unit by unit along dependency edges, \tool first generates comprehensive, structured documentation of the source C codebase through a dedicated \emph{DocGen} module inspired by the hierarchical codebase-documentation ideas of CodeWiki~\cite{hoang2025codewiki}. We adapt this stage for migration by first clustering the repository at the file level, then lifting these groups into higher-level component abstractions, and finally customizing prompts so the generated documentation emphasizes abstract feature that are valuable for Rust generation, rather than merely describing the original C implementation. The resulting documentation acts as a repository-level migration blueprint, capturing module interactions, system organization, data-flow conventions, and architectural rationale, thereby guiding the generation of idiomatic Rust that preserves the structure and design philosophy of the original system instead of its syntactic surface. To ensure semantic completeness, we introduce a documentation-guided iterative refinement mechanism that compares the documentation of the generated Rust codebase against that of the original C codebase, systematically identifying gaps in coverage and structural alignment. Finally, we incorporate execution-aware code revision through test-driven feedback, where dynamic test execution surfaces behavioral inconsistencies that static analysis and documentation comparison alone cannot detect. Our contributions are as follows:

\begin{itemize}[leftmargin=*]
\item We propose a documentation-driven migration paradigm that leverages automatically generated codebase documentation as an intermediate representation for repository-level C-to-Rust translation. By encoding whole-codebase understanding into structured documentation before translation, this approach enables the generation of idiomatic Rust that preserves architectural intent rather than merely replicating syntactic structure.

\item We introduce a documentation-guided iterative refinement mechanism that assesses and improves the generated Rust code by comparing its documentation against that of the original C codebase, promoting semantic completeness and structural alignment across the entire repository.

\item We incorporate execution-aware code revision with test-driven feedback, enabling the system to iteratively correct behavioral inconsistencies through dynamic test execution and achieve functional correctness beyond what static translation can guarantee.

\item We evaluate \tool\ on eight large-scale, real-world C repositories ranging from 11K to 84K LoC, addressing a key limitation of prior work that has primarily focused on small benchmarks and isolated functions. \tool\ compiles every target under both Kimi-K2-Instruct and GPT-5.4, while baseline LLM translators (Self-Repair, EvoC2Rust) fail to compile any repository at this scale. With the open-weight Kimi-K2-Instruct, \tool\ exceeds an agentic Claude Code baseline on feature preservation (\textbf{93.26\%} vs.\ \textbf{52.52\%}) and on cross-evaluation test pass rate (\textbf{95.17\%} vs.\ \textbf{79.85\%}); with GPT-5.4 these figures rise to \textbf{97.76\%} and \textbf{98.70\%}, alongside \textbf{99.41\%} API-level and \textbf{98.47\%} file-level safe rates.

\end{itemize}

%% file: sections/related_work.tex
\section{Related Work}
\label{sec:related-work}

\subsection{Rule-Based C-to-Rust Migration}
Early C-to-Rust systems relied on transpilation and static analysis. C2Rust directly translates syntax but largely preserves C structure and heavy \texttt{unsafe} usage \cite{Emre2021TranslatingCTD, Ling2022InRWE}. Laertes reduces some of this unsafety through borrow-checker-guided pointer rewriting, though only a limited fraction of pointers can be converted and a few failures can keep large alias classes unsafe \cite{Emre2021TranslatingCTD, Emre2023AliasingLOF}. Other tools target specific idiom gaps such as pthread locks, \texttt{FILE*} I/O, output parameters, and unions \cite{Hong2023ConcratAAH, Hong2025ForcratAIP, Hong2024DontWBS, Hong2024ToTOK}. Scylla achieves safe Rust for a restricted C subset \cite{Fromherz2024ScyllaTAB}. These methods provide useful static foundations, but they do not recover repository-level intent or produce consistently idiomatic Rust.

\subsection{LLM-Based Code Translation}
Large language models enable more flexible translation and repair. Some methods rely on intermediate specifications or summaries, including SpecTra and related work on natural-language or formal specification guidance \cite{Nitin2024SpecTraETAH, tai2025nlmiddle, saha2024specdriven}. Others use additional validation signals: VERT uses WebAssembly for candidate checking \cite{Yang2024VERTVEAE}, Syzygy translates tests alongside code and validates through execution \cite{Shetty2024SyzygyDCW}, SACTOR separates semantic preservation from idiomatic refinement \cite{Zhou2025SACTORLCL}, C2SaferRust combines symbolic slicing with LLM repair \cite{Nitin2025C2SaferRustTCG}, and ACToR uses adversarial agent collaboration \cite{Li2025AdversarialACX}. These methods improve local correctness and safety, but they still reason mainly over functions, files, slices, or translated test units rather than a whole-codebase representation.

\subsection{Repository-Level Migration and Benchmarks}
Recent work has started to address repository-level migration. RustMap translates mutually dependent functions with call-graph analysis and bottom-up repair \cite{Cai2025RustMapTPA}. EVOC2RUST, His2Trans, and ENCRUST instead build around a compilable Rust scaffold and iterative refinement \cite{Wang2025EVOC2RUSTASI, Wang2026BuildAwareICC, sim2026encrust}. Other approaches use knowledge graphs for cross-file relations or preprocessing and refactoring before translation \cite{Yuan2025ProjectLevelCTM, Dehghan2025TranslatingLCN}. Beyond migration, RPG and RPG-Encoder study repository-level graph representations for generation and comprehension \cite{Luo2025RPG, Luo2026RPGEncoder}. Benchmarks such as RustRepoTrans, CRUST-Bench, and SWE-bench show that realistic repository tasks remain difficult for current systems \cite{Ou2024RustRepoTransRCH, Khatry2025CRUSTBenchACAS, jimenez2024swebench}. \tool differs by treating documentation as the repository-level representation used for planning, translation, requirement checking, and execution-aware repair.

\subsection{Multi-Agent Coding Frameworks and Repository Tooling}
\tool{} also draws on multi-agent software-engineering systems such as MetaGPT, ChatDev, AgileCoder, and MapCoder, which decompose planning, generation, and review across specialized agents \cite{hong2024metagpt, qian2024chatdev, nguyen2024agilecoder, islam2024mapcoder}. These systems improve long-horizon code generation, but they target green-field development rather than migration of an existing repository. Our framework also depends on repository-level documentation tools: RepoAgent and CodeWiki generate structured hierarchical descriptions of large codebases, which \tool{} repurposes as both a migration input and an evaluation oracle \cite{luo2024repoagent, hoang2025codewiki}. The execution-aware stage further connects to self-debugging from execution feedback \cite{chen2024selfdebug}, and the requirement-comparison stage uses LLM-as-a-judge ideas for documentation equivalence \cite{zheng2023judging}.

%% file: sections/migraagent.tex
\section{The \tool Framework}
\begin{figure*}
    \centering
    \includegraphics[width=1\textwidth]{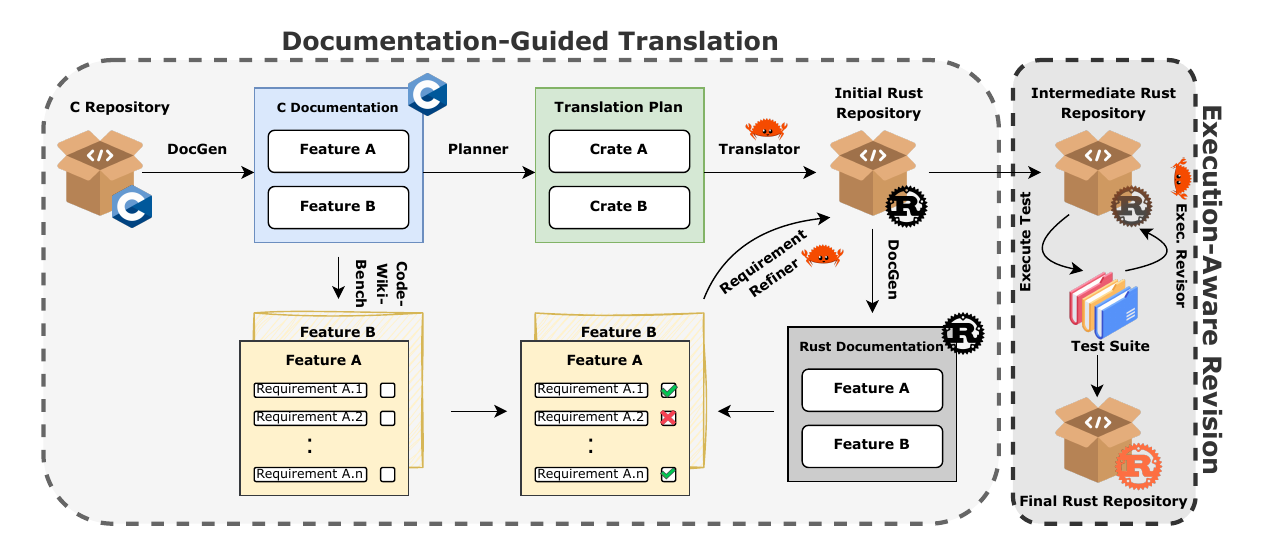}
    \caption{Overview of \tool. DocGen module produces source-side documentation, the Planner turns it into crate-level plans, the Translator generates Rust code, the RequirementRefiner repairs missing functionality by comparing source and target documentation, and the execution-aware stage uses translated tests to fix runtime errors.}
    \label{figs:framework-overview}
\end{figure*}

We introduce \tool, a multi-agent framework for repository-level code migration. The key idea is to use repository-level documentation as the shared representation for planning the translation, refining missing requirements, and repairing execution failures. Figure~\ref{figs:framework-overview} shows the pipeline: generate source-side documentation, translate through crate-level plans, compare source and target documentation to recover missing functionality, and use translated tests to repair runtime errors.

\subsection{Documentation-Guided Code Translation}
\label{doc-guided}

Migration is not a file-by-file rewrite; it requires recovering repository structure and intent. We therefore treat repository-level documentation as the intermediate representation for initial translation. Given a source repository~$S$, we run a dedicated \emph{DocGen} module to produce holistic documentation~$S_{\mathrm{doc}}$ that captures both architecture and component semantics. This module is inspired by repository-documentation systems such as RepoAgent~\cite{luo2024repoagent} and CodeWiki~\cite{hoang2025codewiki}, but it is implemented inside \tool{} and specialized for migration: it clusters code first at the file level, then expands to component-level structure, and prompts for feature-oriented summaries that describe what each subsystem does and how it should be preserved in Rust. We map each high-level feature in $S_{\mathrm{doc}}$ to a Rust crate and use the following agents for planning, implementation, and integration.

\textbf{Planner.}
The \emph{Planner} turns $S_{\mathrm{doc}}$ into a crate-level implementation plan. Using \textit{read\_documentation} and \textit{read\_code\_components}, it resolves architectural and API details from both documentation and source code, then writes an \texttt{IMPLEMENTATION\_PLAN.md} file that specifies the Rust crate structure and component responsibilities.

\textbf{Translator.}
The \emph{Translator} implements each planned crate using \textit{read\_code\_components} and \textit{str\_replace\_editor}. It uses \textit{cargo\_check} to iterate until the crate compiles and \textit{detect\_unsafe} to identify and revise unsafe regions.

\textbf{Synthesizer.}
After per-crate translation, the \emph{Synthesizer} performs repository-level integration. It resolves cross-crate dependencies, aligns interfaces and shared abstractions, and writes a repository \texttt{README.md}.

\subsection{Requirement-Driven Code Refinement}
\label{requirement-refine}

We next use documentation to estimate and improve functional preservation. Our DocGen module generates documentation $T_{\mathrm{doc}}$ for the translated repository~$T$, which we compare against the source documentation $S_{\mathrm{doc}}$. We approximate functional preservation through documentation equivalence:
\begin{equation}
\texttt{CodeEquiv}(S, T) \approx \texttt{DocEquiv}(S_{\mathrm{doc}}, T_{\mathrm{doc}})
\end{equation}

Here, $\texttt{DocEquiv}$ is the CodeWikiBench~\cite{hoang2025codewiki} score for how well $T_{\mathrm{doc}}$ matches $S_{\mathrm{doc}}$, computed with LLM-as-judge protocols~\cite{zheng2023judging}. 

\textbf{RequirementRefiner.} 
When $T_{\mathrm{doc}}$ fails to match $S_{\mathrm{doc}}$, the \emph{RequirementRefiner} edits the translated code with \textit{str\_replace\_editor}, checks compilation with \textit{cargo\_check}, and uses \textit{detect\_unsafe} to handle unsafe code. In practice, this stage recovers omitted features, mismatched APIs, and repository-level behaviors that compile cleanly but are still absent or underspecified in the translated code. It iteratively repairs missing or misaligned functionality.

\subsection{Execution-Aware Code Revision}
\label{execution-aware}
Documentation comparison cannot expose all runtime errors. We therefore add an execution-aware stage that, in the spirit of self-debugging with execution feedback~\cite{chen2024selfdebug}, translates and runs maintainer tests to repair behavioral bugs. We select the requirement-driven refinement version with the highest feature preservation score as the starting point for this execution-aware stage.

\textbf{TestTranslator.} The \emph{TestTranslator} converts the source test suite to the target language and uses \textit{cargo\_test\_no\_run} to ensure inserted tests remain executable.

\textbf{ExecutionRevisor.} The \emph{ExecutionRevisor} analyzes failing tests and updates only the translated code, not the tests. It shares the \emph{RequirementRefiner} toolset and uses \textit{cargo\_single\_test} to debug failing cases one by one. This stage is important because many migration bugs are behavioral rather than documentary: edge cases, state updates, and protocol mismatches may survive documentation alignment but still fail under concrete execution.

Together, these stages mirror human migration: understand the codebase, translate and integrate it, then use tests to repair remaining behavioral bugs.

%% file: sections/evaluation_design.tex
\section{Experimental Setup}

\definecolor{kimicolor}{HTML}{FFFFFF}
\definecolor{gptcolor}{HTML}{CED2EB}
\newcommand{\kimi}[1]{\cellcolor{kimicolor}#1}
\newcommand{\gpt}[1]{\cellcolor{gptcolor}#1}

\subsection{Dataset}

Existing benchmarks for repository-level C-to-Rust migration are limited in scale, with most repositories containing fewer than 1,000 lines of code (LoC), in contrast to broader repository-level coding benchmarks that target realistic, full-project tasks~\cite{jimenez2024swebench}. To enable evaluation in more realistic settings, we manually curate a set of eight C repositories from GitHub, spanning diverse domains and sizes ranging from 11.4K to 83.7K LoC. Each repository is required to include a \texttt{tests/} directory with ground-truth test cases, which serve as the basis for evaluation in our execution-aware setting.

\subsection{Metrics}
To assess the translated codebase, we use four dimensions: \textit{Project Compilability}, \textit{Feature Preservation}, \textit{Functional Correctness}, and \textit{Safety}.

\noindent{\textbf{Project Compilability.}}
A repository is considered compilable if \texttt{cargo check} completes without errors.

\noindent{\textbf{Feature Preservation}}
For repositories that compile successfully, we compare source and translated documentation with CodeWikiBench to measure how well the translated code preserves the source feature surface. Each rubric is organized as a tree in which high-level components are decomposed into leaf-level requirements that capture concrete functional and architectural elements. The final feature preservation score is computed as:

\begin{equation}
\textit{FCV} = \frac{\sum_{\ell \in \mathcal{L}} w_{\ell} \cdot \mathbf{1}(\ell)}{\sum_{\ell \in \mathcal{L}} w_{\ell}},
\end{equation}

where $\mathcal{L}$ denotes the set of leaf-level requirements, $w_{\ell}$ is the importance weight of requirement $\ell$, and $\mathbf{1}(\ell)$ indicates whether requirement $\ell$ is preserved.

\noindent{\textbf{Functional Correctness.}}
Feature preservation alone does not guarantee correct execution. We therefore evaluate translated repositories with test suites produced by both \tool and Claude Code, and we report \emph{Test Pass Rate (TPR)} as the percentage of translated tests that pass during execution.

\noindent{\textbf{Safety.}}
We report \emph{Safe Rate (SR)} at two granularities: \emph{SR (A)} measures the fraction of public APIs that do not require \texttt{unsafe}, and \emph{SR (F)} measures the fraction of files that contain no \texttt{unsafe} blocks.


\subsection{Baselines.}
For traditional baselines, we use C2Rust~\cite{c2rust}, as it is widely recognized and adopted by the community for C-to-Rust migration. For LLM-based approaches, we consider Self-Repair~\cite{c2rust-bench} and EvoC2Rust~\cite{Wang2025EVOC2RUSTASI}, which represent recent efforts in leveraging large language models for code translation. In addition, to evaluate the competitiveness and practicality of \tool, we include Claude Code~\cite{santos2025claudecode} as a representative agentic baseline.

\subsection{Models}
To enable scalability analysis, we evaluate our approach using both an open-source model (\texttt{Kimi-K2-Instruct}~\cite{kimi2025k2}) and a proprietary model (\texttt{GPT-5.4}~\cite{openai2025gpt5}), representing competitive capabilities across diverse deployment settings.

%% file: sections/experiment.tex
\section{Evaluation}

This section addresses three questions: (1) \textit{Can current methods translate
repository-scale C codebases into compilable Rust} (Sec.~\ref{sec:eval-compile})?,
(2) \textit{If so, do the translated repositories preserve source functionality and execute
it correctly} (Sec.~\ref{sec:eval-feature}, Sec.~\ref{sec:eval-correctness})?
\textit{And do they retain Rust's safety benefits (Sec.~\ref{sec:eval-safety})?}

\subsection{Project Compilability}
\label{sec:eval-compile}

\begin{table}[h]
\centering
\scriptsize
\setlength{\tabcolsep}{3pt}
\renewcommand{\arraystretch}{1.4}
\setlength{\aboverulesep}{0.5pt}
\setlength{\belowrulesep}{0.5pt}

\resizebox{\textwidth}{!}{
\begin{tabular}{lccccccccc}
\toprule
\multirow{2}{*}{\textbf{Repo}}
& \multirow{2}{*}{\textbf{\#LoC}}
& \multirow{2}{*}{\textbf{C2Rust}}
& \multicolumn{3}{c}{\textbf{Kimi-K2-Instruct}}
& \multicolumn{3}{c}{\textbf{GPT-5.4}}
& \multirow{2}{*}{\textbf{Claude Code}} \\
\cmidrule(lr){4-6}\cmidrule(lr){7-9}
& & 
& Self-Repair & EvoC2Rust & \tool (Ours)
& Self-Repair & EvoC2Rust & \tool (Ours)
& \\
\midrule
\textbf{libplist}   & 17.6K & \checkyes & \crossno & \crossno & \checkyes & \crossno & \crossno & \checkyes & \checkyes \\
\textbf{check}      & 20.3K & \checkyes & \crossno & \crossno & \checkyes & \crossno & \crossno & \checkyes & \checkyes \\
\textbf{stb}        & 83.7K & \checkyes & \crossno & \crossno & \checkyes & \crossno & \crossno & \checkyes & \checkyes \\
\textbf{klib}       & 12.5K & \checkyes & \crossno & \crossno & \checkyes & \crossno & \crossno & \checkyes & \checkyes \\
\textbf{libcbor}    & 13.9K & \checkyes & \crossno & \crossno & \checkyes & \crossno & \crossno & \checkyes & \checkyes \\
\textbf{Monocypher} & 13.3K & \checkyes & \crossno & \crossno & \checkyes & \crossno & \crossno & \checkyes & \checkyes \\
\textbf{libfixmath} & 15.9K & \checkyes & \crossno & \crossno & \checkyes & \crossno & \crossno & \checkyes & \checkyes \\
\textbf{libyaml}    & 11.4K & \checkyes & \crossno & \crossno & \checkyes & \crossno & \crossno & \checkyes & \checkyes \\
\bottomrule
\end{tabular}
}
\captionsetup{skip=10pt}
\caption{Repository-level compilation success across translation methods and
  model backbones. \checkyes\ indicates a fully buildable Cargo project;
  \crossno\ indicates the repository fails to compile end-to-end.}
\label{tab:compilability}
\end{table}

Table~\ref{tab:compilability} reports compilation outcomes
across all eight benchmark repositories under both model backbones.
\tool consistently produces fully buildable Cargo projects across all repositories (11K-83K LoC). This result highlights the effectiveness of \tool’s documentation-guided planning combined with its iterative refinement loop. By equipping agents with the \textit{cargo\_check} tool to validate every code change on-the-fly, the system ensures that modifications preserve compilability and avoid cascading errors - closely replicating how human developers write, test, and refine code incrementally. The agentic baseline Claude Code also achieves consistent compilation success across all repositories, confirming the value of multi-turn agentic workflows for repository-level migration. C2Rust likewise produces buildable output in every case; however, as will be discussed in Section~\ref{sec:eval-compile}, it achieves this by performing direct syntactic transpilation that retains the vast majority of C-style constructs, resulting in heavily unsafe Rust code. In contrast, Self-Repair and EvoC2Rust fail to produce end-to-end compilable repositories under both Kimi-K2-Instruct and GPT-5.4, highlighting their limitations in handling complex cross-module dependencies and global structural consistency at repository scale.

\subsection{Feature Preservation}
\label{sec:eval-feature}

\begin{figure}[h]
    \centering

    \begin{subfigure}{0.48\linewidth}
        \centering
        \includegraphics[width=\linewidth]{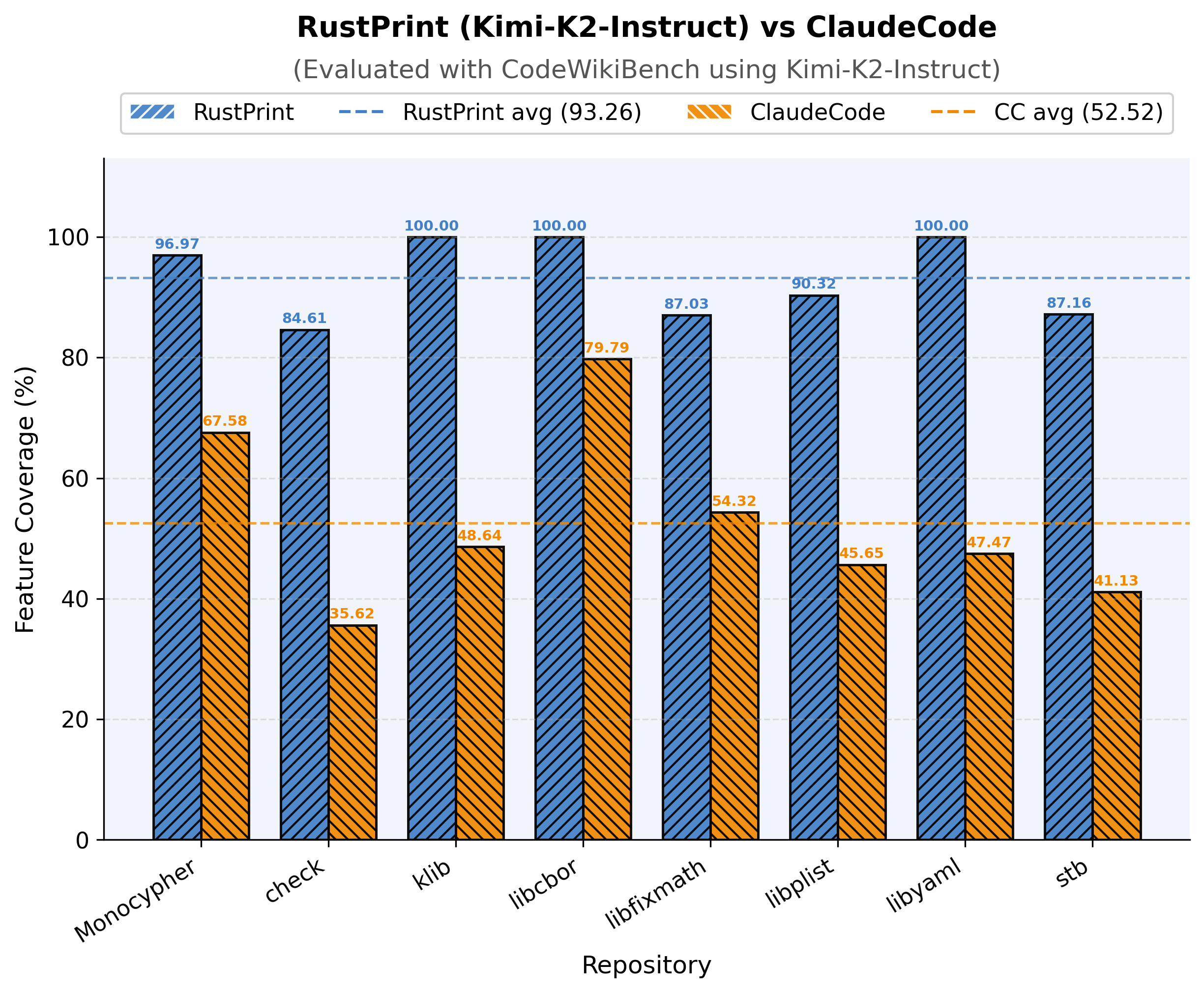}
        \caption{}
        \label{fig:feature-kimi}
    \end{subfigure}
    \hfill
    \begin{subfigure}{0.48\linewidth}
        \centering
        \includegraphics[width=\linewidth]{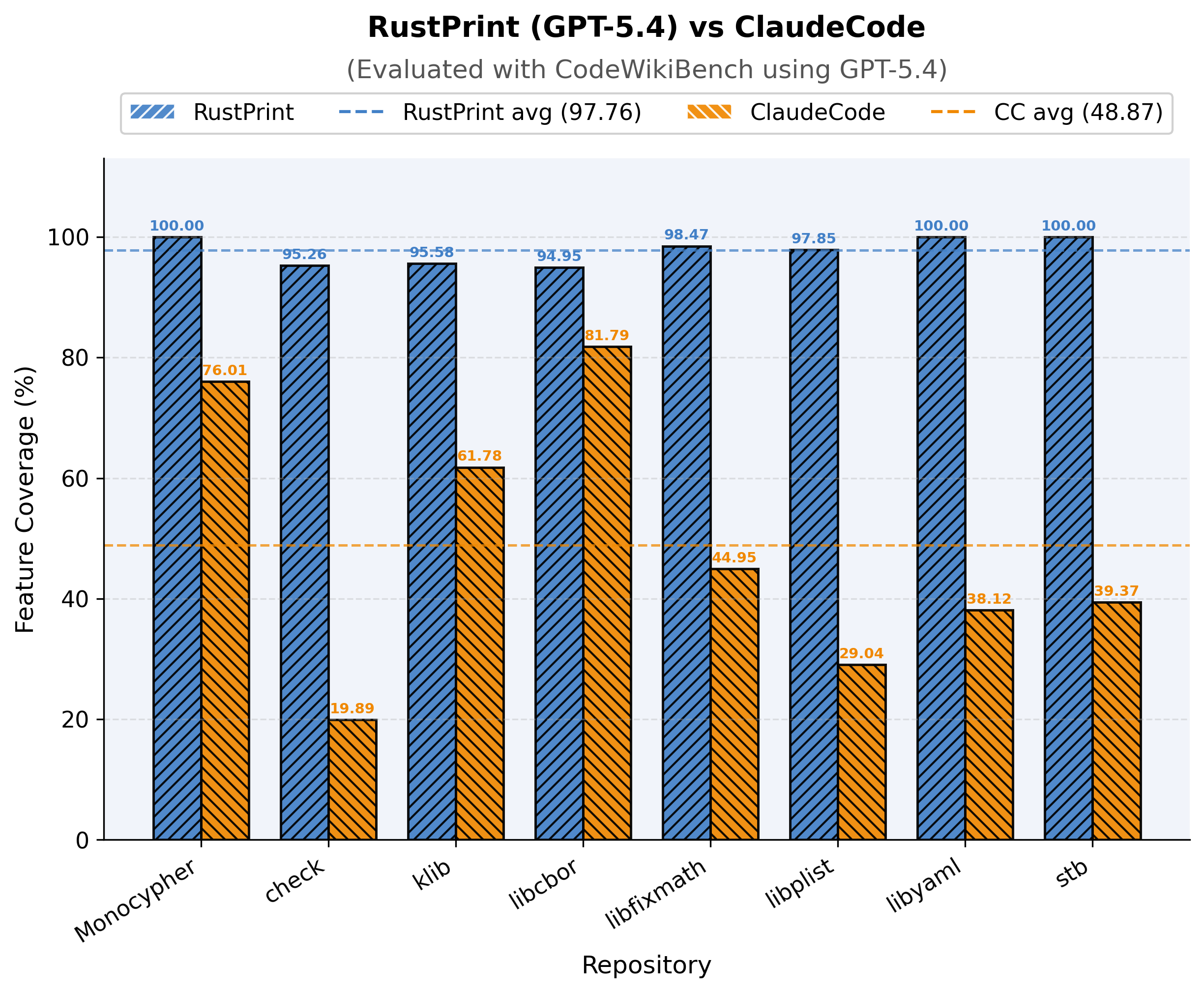}
        \caption{}
        \label{fig:feature-gpt}
    \end{subfigure}

    \caption{Per-repository feature preservation scores (\%), comparing \tool\ to ClaudeCode under different model backbones. \textbf{(a)} Kimi-K2-Instruct: \tool\ achieves \textbf{93.26\%} versus ClaudeCode's \textbf{52.52\%}. \textbf{(b)} GPT-5.4: \tool\ reaches \textbf{97.76\%} versus ClaudeCode's \textbf{48.87\%}.}
    
    \label{fig:feature-preservation}
\end{figure}

\begin{figure}[h]
    \centering

    \begin{subfigure}{0.48\linewidth}
        \centering
        \includegraphics[width=\linewidth]{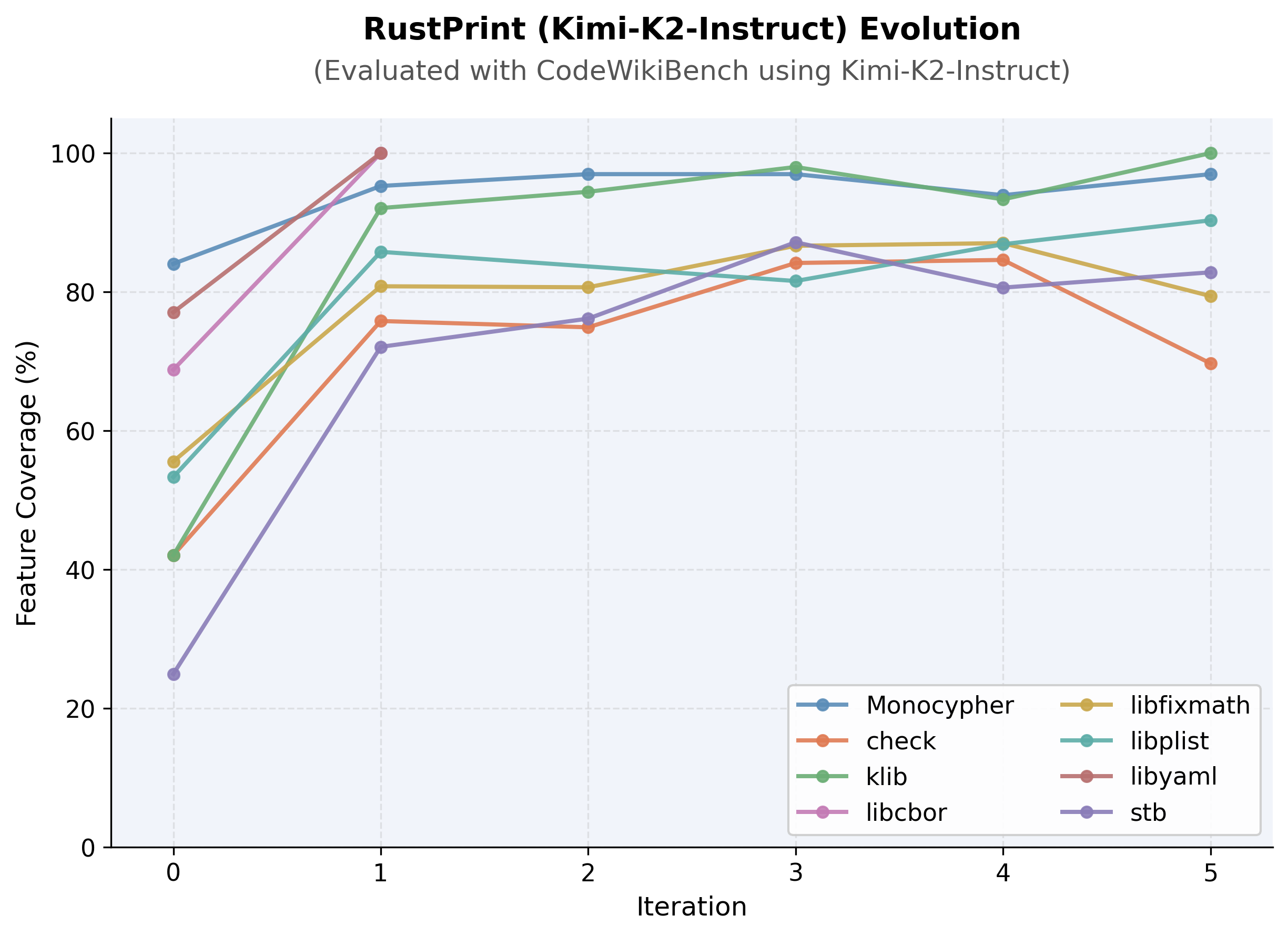}
        \caption{}
        \label{fig:feature-evo-a}
    \end{subfigure}
    \hfill
    \begin{subfigure}{0.48\linewidth}
        \centering
        \includegraphics[width=\linewidth]{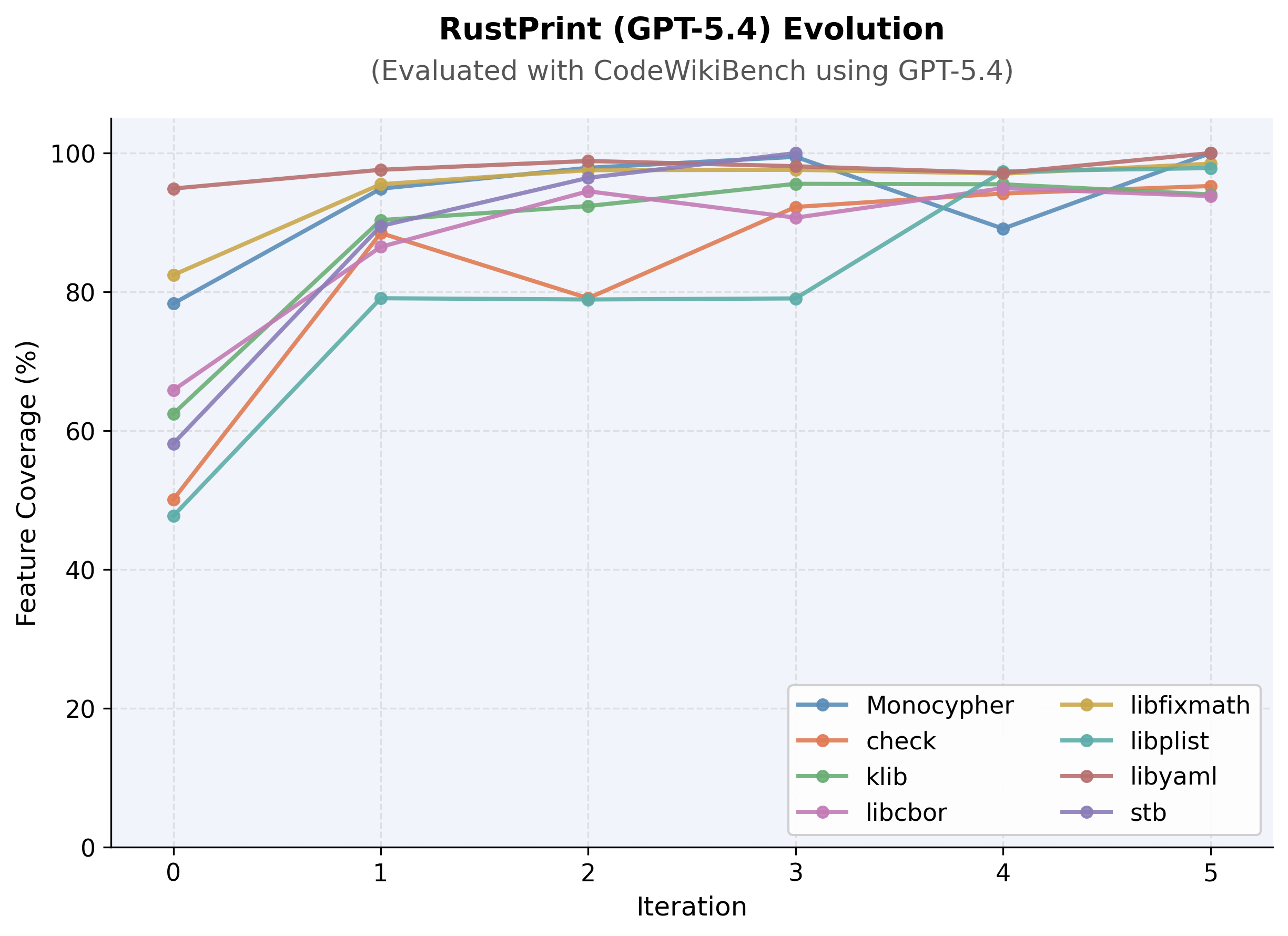}
        \caption{}
        \label{fig:feature-evo-b}
    \end{subfigure}

    \caption{Feature preservation across refinement iterations (0--5) for
    \tool\ on the eight benchmark repositories.
    (\textbf{a}) Kimi-K2-Instruct. 
    (\textbf{b}) GPT-5.4.
    Both backbones show sharp gains within the first one or two iterations, 
    after which scores stabilise near completion.}

    \label{fig:feature-evolution}
\end{figure}

Compilability is necessary but not sufficient: a buildable Rust project that
omits or alters source functionality is not a faithful translation. To probe
\emph{feature preservation}, we evaluate the translated repositories against
\textbf{CodeWikiBench}, which scores how completely the translated code
reproduces the documented functional surface of the source. We report two
complementary views: a final-state comparison against the Claude Code baseline
(Fig.~\ref{fig:feature-preservation}), and the per-iteration evolution
that reveals the dynamics of refinement (Fig.~\ref{fig:feature-evolution}).
We restrict this comparison to \tool\ and Claude Code: C2Rust is direct unsafe
transpilation, and Self-Repair and EvoC2Rust do not produce compilable
repositories at this scale.

\textbf{Main Result.} Figure~\ref{fig:feature-preservation} shows the Feature Coverage Score (\textit{FCV}) scores after iterative refinement for both methods across all eight repositories. \tool significantly outperforms Claude Code, surpassing it by more than \textbf{40\%} on \textit{FCV} when evaluating the generated codebase documentation on CodeWikiBench. In particular, \tool attains an average \textit{FCV} of \textbf{93.26\%} with Kimi-K2-Instruct and \textbf{97.76\%} with GPT-5.4. In contrast, Claude Code, despite consistently producing compilable code, only reaches an average of \textbf{52.52\%} (under Kimi-K2-Instruct evaluation) and \textbf{48.87\%} (under GPT-5.4 evaluation). This substantial gap highlights the critical importance of moving beyond direct code-to-code translation. By treating comprehensive documentation as the central migration blueprint, \tool enables agents to better understand and faithfully preserve functionality. 

\textbf{Evolution Across Iterations.} Figure~\ref{fig:feature-evolution} further illustrates the effectiveness of \tool’s requirement-driven refinement process. Across both Kimi-K2-Instruct and GPT-5.4, the \textit{FCV} shows a strong overall upward trend across refinement iterations. This progression validates our documentation-guided iterative mechanism, which enables agents to progressively recover missing requirements and enhance semantic alignment by leveraging mismatches between the source and generated documentation as feedback to refine the translated codebase. Moreover, this process closely mimics the human workflow in which developers iteratively compare their implementation against specifications or documentation to identify and address gaps.

\subsection{Functional Correctness}
\label{sec:eval-correctness}


We further evaluate the functional correctness of the translated repositories by comparing \tool with Claude Code on test suites. We first leverage the \textit{TestTranslator} agent to generate test suites for both methods. However, we observe that the \textit{TestTranslator} of each system tends to skip or omit tests corresponding to features that its own translated repository failed to preserve. This behavior introduces a self-alignment bias when a method is evaluated solely on its own generated tests. To enable a more rigorous and fair comparison, we introduce a cross-test evaluation protocol. We employ an additional agent (backed by GPT-5.4) equipped with a \textit{copy\_test} tool. This agent systematically traverses every test function from one translated repository (e.g., generated by \tool) and carefully adapts it to the other repository (e.g., generated by Claude Code). During adaptation, the agent preserves the original test logic and intent while appropriately modifying API calls, data structures, and interface surfaces to match the target codebase. This cross-test adaptation process allows us to evaluate each translated artifact under both its own test suite and the independent test suite produced by the other system, providing a more objective measure of true functional correctness.

\begin{table}[h]
\centering
\scriptsize
\setlength{\tabcolsep}{5pt}
\renewcommand{\arraystretch}{1.05}
\resizebox{0.80\textwidth}{!}{
\begin{tabular}{l>{\columncolor{rowgraylight}}c c >{\columncolor{rowgraylight}}c c >{\columncolor{rowgraylight}}c c}
\toprule
\textbf{Repo}
& \multicolumn{2}{c}{\makecell{\textbf{\tool}\\[-1pt]\scriptsize Kimi-K2-Instruct}}
& \multicolumn{2}{c}{\makecell{\textbf{\tool}\\[-1pt]\scriptsize GPT-5.4}}
& \multicolumn{2}{c}{\makecell{\textbf{Claude Code}}} \\
\cmidrule(l){2-3}\cmidrule(l){4-5}\cmidrule(l){6-7}
& TPR\,(R) & TPR\,(C)
& TPR\,(R) & TPR\,(C)
& TPR\,(R) & TPR\,(C) \\
\midrule
\textbf{libplist}   & 100.00 & 100.00 & 100.00 & 100.00 & 73.33 & 85.19 \\
\textbf{check}      & 98.00  & 100.00 & 100.00 & 100.00 & 90.50 & 100.00 \\
\textbf{stb}        & 100.00 & 100.00 & 100.00 & 100.00 & 35.71 & 84.62 \\
\textbf{klib}       & 97.37  & 98.08  & 100.00 & 100.00 & 73.08 & 98.08 \\
\textbf{libcbor}    & 98.82  & 100.00 & 100.00 & 100.00 & 84.62 & 100.00 \\
\textbf{Monocypher} & 72.73  & 95.83  & 93.18  & 97.91  & 65.91 & 91.67 \\
\textbf{libfixmath} & 94.74  & 100.00 & 95.77  & 100.00 & 45.07 & 100.00 \\
\textbf{libyaml}    & 75.71  & 91.42  & 95.16  & 97.14  & 61.29 & 88.57 \\
\bottomrule
\end{tabular}
}
\captionsetup{skip=10pt}
\caption{Cross-evaluation of test pass rate (TPR, \%) under independently
  generated test suites. \textbf{R}: tests authored by \tool;
  \textbf{C}: tests authored by Claude Code. Each translated artefact is
  evaluated under both suites; high TPR on the \emph{other} method's suite
  indicates genuine functional correctness rather than self-test alignment.}
\label{tab:tpr-cross}
\end{table}

Table~\ref{tab:tpr-cross} presents the results of this cross-evaluation. As shown in the table, \tool achieves strong functional correctness across both Kimi-K2-Instruct and GPT-5.4. It consistently attains high TPR scores on most repositories under both its own generated tests (\textbf{R}) and Claude Code’s tests (\textbf{C}). In contrast, Claude Code shows noticeably lower test pass rates overall. While it performs better when evaluated on its own test suite (\textbf{C}), its performance drops considerably when tested against \tool-generated test suites (\textbf{R}). Notable drops in TPR,(R) can be observed on repositories such as \texttt{stb} (35.71\%), \texttt{libfixmath} (45.07\%), and \texttt{libyaml} (61.29\%). These results are consistent with the feature preservation findings in Section~\ref{sec:eval-feature}, reinforcing that the performance gap stems from differences in preserved functionality.

Aggregating across all 16 (repository\,$\times$\,test-suite) cells, \tool\ averages \textbf{98.70\%} TPR under GPT-5.4 and \textbf{95.17\%} under Kimi-K2-Instruct, versus \textbf{79.85\%} for Claude Code. Even with the open-weight Kimi-K2-Instruct backbone, \tool\ exceeds Claude Code on aggregate TPR by about \textbf{15 percentage points}.



\subsection{Safety Assessment}
\label{sec:eval-safety}

\begin{figure}[h]
\centering

\begin{minipage}{0.48\textwidth}
\centering
\scriptsize
\setlength{\tabcolsep}{5pt}
\renewcommand{\arraystretch}{1.05}

\resizebox{\linewidth}{!}{
\begin{tabular}{lcc}
\toprule
\textbf{Method} & \textbf{SafeRate (A)} & \textbf{SafeRate (F)} \\
\midrule
\textbf{C2Rust} & 0.00 & 0.00 \\
\textbf{Claude Code} & 99.09 & 91.10 \\
\cmidrule(lr){1-3}
\textbf{EvoC2Rust$_{\scriptsize \text{Kimi-K2-Instruct}}$} & 94.79 & \textbf{96.60} \\
\textbf{Self-Repair$_{\scriptsize \text{Kimi-K2-Instruct}}$} & \underline{95.72} & 84.92 \\
\textbf{\tool$_{\scriptsize \text{Kimi-K2-Instruct}}$} & \textbf{96.23} & \underline{96.19} \\
\cmidrule(lr){1-3}
\textbf{EvoC2Rust$_{\scriptsize \text{GPT-5.4}}$} & 95.13 & 96.74 \\
\textbf{Self-Repair$_{\scriptsize \text{GPT-5.4}}$} & \underline{99.19} & \underline{97.71} \\
\textbf{\tool$_{\scriptsize \text{GPT-5.4}}$} & \textbf{99.41} & \textbf{98.47} \\
\bottomrule
\end{tabular}
}
\caption{SafeRate (A) and SafeRate (F) across translation methods and model backbones.}
\label{tab:safetable}
\end{minipage}
\hfill
\begin{minipage}{0.48\textwidth}
\centering
\includegraphics[width=\linewidth]{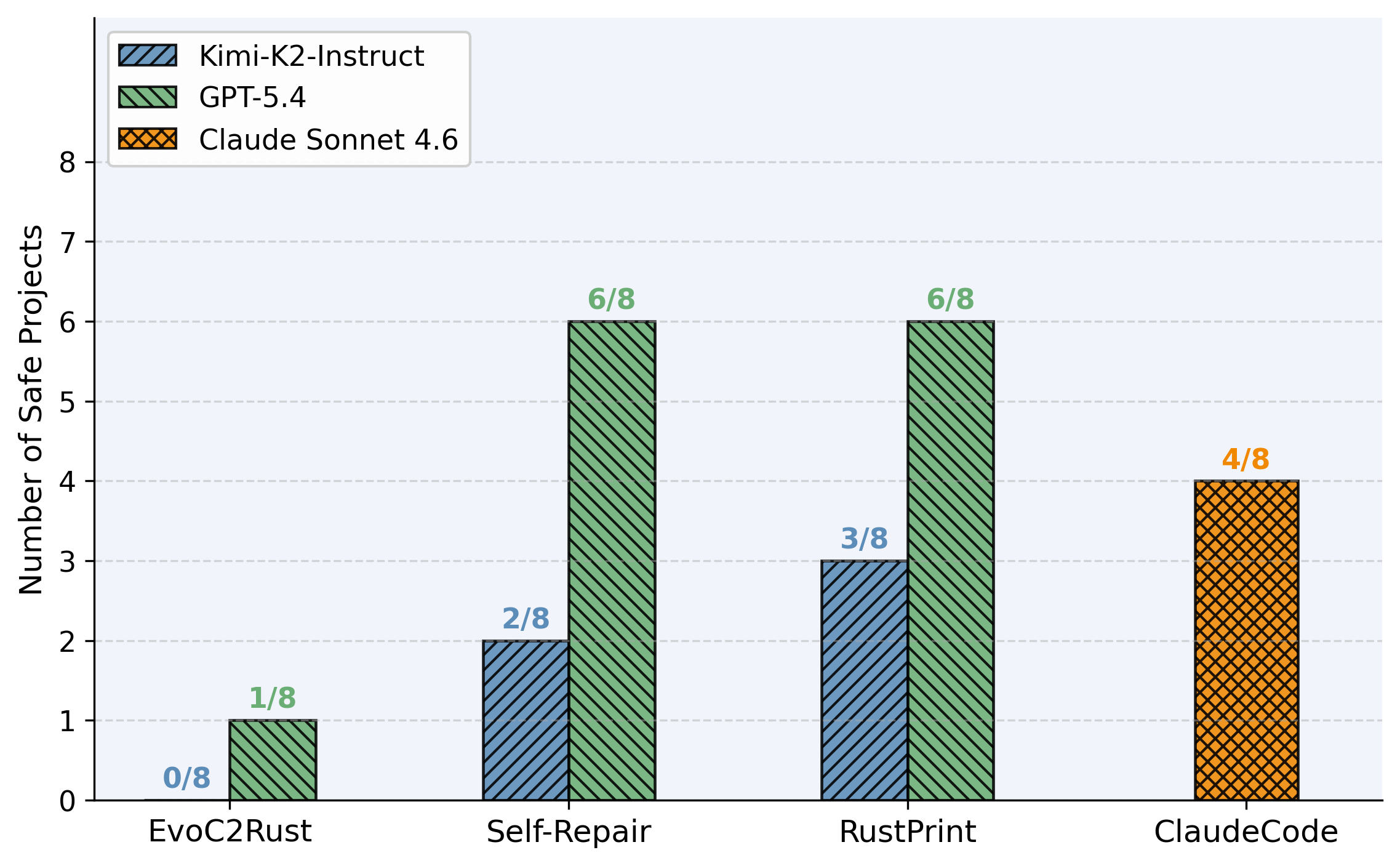}
\caption{Number of fully safe projects per method and backbone.}
\label{fig:safefigure}
\end{minipage}

\end{figure}

Figures~\ref{tab:safetable} and~\ref{fig:safefigure} summarize the safety
performance across methods and model backbones. With the commercial Claude Code
system, the translated repositories achieve a competitive API-level safe rate
of \textbf{99.09\%} and a file-level safe rate of \textbf{91.10\%}. \tool\
also demonstrates strong safety characteristics. With GPT-5.4, it achieves the
highest scores among all methods, reaching \textbf{99.41\%} API-level and
\textbf{98.47\%} file-level safe rates, compared to \textbf{95.13\%}/
\textbf{96.74\%} for EvoC2Rust and \textbf{99.19\%}/\textbf{97.71\%} for
Self-Repair. Under Kimi-K2-Instruct, \tool\ remains competitive with
\textbf{96.23\%} SafeRate (A) and \textbf{96.19\%} SafeRate (F). In contrast to
direct code-to-code baselines, \tool\ follows repository-level documentation as
the migration blueprint, which reduces pressure to introduce unnecessary
\texttt{unsafe} operations during translation and repair.

These summary views complement the per-project safety evidence. C2Rust keeps
repositories buildable but leaves none fully safe, while Claude Code preserves
high API safety yet still leaves more file-level \texttt{unsafe} residue.
\tool, especially with GPT-5.4, is the only approach that combines strong
compilation results with very high SafeRate scores and the largest number of
fully safe translated repositories.

%% file: sections/limitations.tex
The results support three main takeaways about repository-level migration. First, planning and refinement matter more than one-shot translation at this scale: \tool compiles all eight repositories under both Kimi-K2-Instruct and GPT-5.4, while the simpler LLM baselines fail to produce a single compilable repository. This suggests that repository migration is less a raw code-generation problem and more a coordination problem across files, interfaces, and repair steps. Second, documentation is useful not just as background context but as an operational guide for refinement. The feature-preservation results show large gains over Claude Code, and the refinement curves indicate that much of that gain appears in the first one or two documentation-guided iterations. Third, the cross-suite test results are important because they show that these gains are not only documentary. \tool remains strong on tests written by another method, which indicates that the translated repositories preserve behavior beyond the particular tests generated inside the pipeline.

At the same time, the current study should be read as a strong first result rather than a complete map of the space. The benchmark covers realistic repositories, but broader validation on heavier FFI usage, less standardized build pipelines, and more concurrency-heavy systems would strengthen the claim further. Documentation comparison is also best understood as one useful signal among several: it works especially well when paired with compilation and translated tests, but future versions should combine it with stronger runtime and static checks for critical behaviors. A broader evaluation should also report time, tool usage, and model cost, since those factors matter in practice alongside quality. Framed this way, the present results show that documentation-guided migration is already viable on realistic repositories while leaving clear and actionable paths for improving both the method and its evaluation.

%% file: sections/conclusions.tex
\section{Conclusions \& Future Work}

This paper presented \tool, a documentation-guided multi-agent framework for repository-level C-to-Rust migration. Rather than treating migration as one-shot code generation, \tool decomposes it into planning, translation, test revision, and repair stages grounded in repository documentation and execution feedback. Documentation serves as the migration blueprint, linking recovered requirements to implementation and validation across files. This design more closely matches how developers handle large migrations with build constraints and cross-file dependencies.

Across eight repositories, \tool outperforms strong baselines in compilability, cross-suite test pass rate, feature preservation, and safety, with especially strong results under GPT-5.4. Taken together, these results show that documentation-guided coordination is a practical way to scale LLM-based translation from isolated files to repository-level engineering tasks.

Future work should test this framework on mixed-language repositories, deeper external dependencies, and build systems that are less standardized than Cargo. It would also be valuable to combine documentation-guided planning with stronger verification signals, including differential testing, property-based testing, and static analysis for safety-critical modules. These extensions would clarify how far the approach generalizes beyond the current C-to-Rust setting.

%% file: sections/appendix.tex
%

\begin{center}
{\bf{\LARGE{Appendices}}}
\end{center}

\vspace{0.5em}

{\hypersetup{linkcolor=black}
\noindent\textbf{\large Contents}\par\medskip
\startcontents[appendices]
\printcontents[appendices]{l}{1}{\setcounter{tocdepth}{2}}
}

\vspace{1em}

\newtcolorbox{promptbox}[1][]{%
  enhanced, breakable,
  colback=white, colframe=black,
  colbacktitle=black, coltitle=white,
  fonttitle=\bfseries,
  boxrule=0.6pt, arc=2pt,
  left=6pt, right=6pt, top=4pt, bottom=4pt,
  title=#1
}

\IfFileExists{sections/appendix/A_algorithm.tex}{\input{sections/appendix/A_algorithm}}{}
\IfFileExists{sections/appendix/B1_implementation.tex}{\input{sections/appendix/B1_implementation}}{}
\IfFileExists{sections/appendix/B2_tools_cargo.tex}{\input{sections/appendix/B2_tools_cargo}}{}
\IfFileExists{sections/appendix/B3_tools_read.tex}{\input{sections/appendix/B3_tools_read}}{}
\IfFileExists{sections/appendix/B4_tools_misc.tex}{\input{sections/appendix/B4_tools_misc}}{}
\IfFileExists{sections/appendix/B5_prompts_plan_trans.tex}{\input{sections/appendix/B5_prompts_plan_trans}}{}
\IfFileExists{sections/appendix/B6_prompts_synth_refine.tex}{\input{sections/appendix/B6_prompts_synth_refine}}{}
\IfFileExists{sections/appendix/B7_prompts_test_exec.tex}{\input{sections/appendix/B7_prompts_test_exec}}{}
\IfFileExists{sections/appendix/C_case_study.tex}{\input{sections/appendix/C_case_study}}{}
\IfFileExists{sections/appendix/D_cost_analysis.tex}{\input{sections/appendix/D_cost_analysis}}{}
\IfFileExists{sections/appendix/EF_limitations_broader.tex}{\input{sections/appendix/EF_limitations_broader}}{}

\stopcontents[appendices]

%% file: sections/appendix/A_algorithm.tex
\newpage
\section{Algorithm}
\label{sec:appendix-algorithm}

\paragraph{Overview.} \tool{} executes a five-stage pipeline: (i) holistic C
documentation generation, (ii) per-crate planning and translation with
compile-safety loops, (iii) workspace-level synthesis, (iv) up to $K$ rounds
of documentation-driven requirement refinement followed by best-version
selection, and (v) up to $L$ rounds of execution-aware revision against
translated tests.
In our experiments we set $K = L = 5$.
Algorithm~\ref{alg:rustprint} formalises this end-to-end procedure.

\begin{algorithm}[H]
\DontPrintSemicolon
\SetKwInOut{Input}{Input}
\SetKwInOut{Output}{Output}
\Input{Source C repository $S$; max refinement rounds $K$; max revision rounds $L$.}
\Output{Translated Rust workspace $T^{\star}_{\mathrm{exec}}$.}
\BlankLine

\tcp{Stage 1: Documentation generation}
$S_{\mathrm{doc}} \leftarrow \textsc{DocGen}(S)$ \;
$\mathcal{C} \leftarrow \textsc{Crates}(S_{\mathrm{doc}})$ \tcp*{one crate per top-level feature}
\BlankLine

\tcp{Stage 2: Per-crate planning and translation}
\ForEach{$c \in \mathcal{C}$}{
  $\pi_c \leftarrow \textsc{Planner}(c, S_{\mathrm{doc}}, S)$ \;
  $T_c \leftarrow \textsc{Translator}(\pi_c, S)$ \;
  \While{\upshape\textsc{cargo\_check}$(T_c) \neq$ \textnormal{``Done''}  \textbf{or}  \textsc{unsafe\_detect}$(T_c) \neq \emptyset$}{
    $T_c \leftarrow$ fix compile errors and minimise unsafe blocks\;
  }
}
\BlankLine

\tcp{Stage 3: Workspace synthesis}
$T^{(0)} \leftarrow \textsc{Synthesizer}(\{T_c\}_{c \in \mathcal{C}})$ \tcp*{resolve cross-crate dependencies}
\BlankLine

\tcp{Stage 4: Requirement-driven refinement}
$\mathcal{V} \leftarrow \emptyset$ \tcp*{candidate versions scored by feature preservation}
\For{$i \leftarrow 0$ \KwTo $K$}{
  $T_{\mathrm{doc}}^{(i)} \leftarrow \textsc{DocGen}(T^{(i)})$ \;
  $(s_i, \mathcal{M}_i) \leftarrow \textsc{DocGenBench}(T_{\mathrm{doc}}^{(i)}, S_{\mathrm{doc}})$ \tcp*{FCV score and failing rubrics}
  $\mathcal{V} \leftarrow \mathcal{V} \cup \{(i, T^{(i)}, s_i)\}$ \;
  \lIf{$i = K$ \textbf{or} $\mathcal{M}_i = \emptyset$}{\textbf{break}}
  $T^{(i+1)} \leftarrow T^{(i)}$ \tcp*{copy current version before editing}
  \ForEach{$r \in \mathcal{M}_i$}{
    $T^{(i+1)} \leftarrow \textsc{RequirementRefiner}(T^{(i+1)}, r)$ \;
  }
}
$T^{\star} \leftarrow U \text{ where } (i, U, s) = \arg\max_{(i, U, s) \in \mathcal{V}} s$ \tcp*{best requirement-refined version}
\BlankLine

\tcp{Stage 5: Execution-aware revision}
$T^{(0)}_{\mathrm{exec}} \leftarrow \textsc{TestTranslator}(S, T^{\star})$ \tcp*{port C tests to selected workspace}
$T^{\star}_{\mathrm{exec}} \leftarrow T^{(0)}_{\mathrm{exec}}$ \;
\For{$j \leftarrow 1$ \KwTo $L$}{
  $\mathcal{F}_{j-1} \leftarrow \{t \mid t \text{ fails in } \textsc{CargoTest}(T^{(j-1)}_{\mathrm{exec}})\}$ \;
  \lIf{$\mathcal{F}_{j-1} = \emptyset$}{\textbf{break}}
  $T^{(j)}_{\mathrm{exec}} \leftarrow T^{(j-1)}_{\mathrm{exec}}$ \tcp*{copy current execution version before editing}
  \ForEach{$t \in \mathcal{F}_{j-1}$}{
    $T^{(j)}_{\mathrm{exec}} \leftarrow \textsc{ExecutionRevisor}(T^{(j)}_{\mathrm{exec}}, t)$ \tcp*{fix using stdout/stderr}
  }
  $T^{\star}_{\mathrm{exec}} \leftarrow T^{(j)}_{\mathrm{exec}}$ \;
}
\BlankLine

\Return{$T^{\star}_{\mathrm{exec}}$}\;
\caption{The end-to-end \tool{} pipeline.}
\label{alg:rustprint}
\end{algorithm}

%% file: sections/appendix/B1_implementation.tex
\newpage
\section{Implementation Details}
\label{sec:appendix-impl}

%% file: sections/appendix/B2_tools_cargo.tex
\subsection{Tool Design}\label{sec:appendix-tools}

Each agent in \tool{} is equipped with a small set of read- and write-tools that
together let it inspect the C source, edit the Rust workspace, and validate its
work. This appendix documents every tool: its purpose, its function signature,
the core code, and the agents that register it.
Table~\ref{tab:appendix-agent-tool-matrix} summarises which agent uses which
tool.

\begin{table}[h]
\centering
\caption{Agent-to-tool registration matrix.}
\label{tab:appendix-agent-tool-matrix}
\resizebox{\textwidth}{!}{%
\begin{tabular}{l c c c c c c}
\toprule
Tool & Planner & Translator & Synthesizer & TestTranslator & RequirementRefiner & ExecutionRevisor \\
\midrule
\code{cargo\_check}          & \crossno & \checkyes & \checkyes & \crossno & \checkyes & \checkyes \\
\code{cargo\_single\_test}   & \crossno & \crossno & \crossno & \crossno & \crossno & \checkyes \\
\code{cargo\_test\_no\_run}  & \crossno & \crossno & \crossno & \checkyes & \crossno & \checkyes \\
\code{copy\_test}            & \crossno & \crossno & \crossno & \checkyes & \crossno & \crossno \\
\code{find\_code\_component} & \checkyes & \checkyes & \checkyes & \checkyes & \checkyes & \checkyes \\
\code{read\_documentation}   & \checkyes & \checkyes & \crossno & \crossno & \checkyes & \crossno \\
\code{str\_replace\_editor}  & \checkyes & \checkyes & \checkyes & \checkyes & \checkyes & \checkyes \\
\code{unsafe\_detect}        & \crossno & \checkyes & \crossno & \crossno & \checkyes & \crossno \\
\bottomrule
\end{tabular}%
}
\end{table}

\newpage
\subsubsection{\code{cargo\_check}}

The \code{cargo\_check} tool runs \texttt{cargo check} inside the translated
Rust repository and returns either a success signal or the compiler's
diagnostic output for the agent to act on. It supports two scopes:
\code{"crate"} (the default, which executes the check inside the current crate
directory and is used during per-module translation) and \code{"workspace"}
(which executes the check at the workspace root and is reserved for synthesis
and single-repo refinement). Persistent attempt counters on the dependency
container let the agent track how many iterations a repair loop has consumed.

\begin{lstlisting}[style=pythonstyle, numbers=left]
async def cargo_check(
    ctx: RunContext[DepsWithRustPath],
    scope: Literal["crate", "workspace"] = "crate",
) -> str:
    cmd = ["cargo", "check"]
    result = subprocess.run(
        cmd,
        cwd=cwd,
        capture_output=True,
        text=True,
        timeout=300,
    )
    if result.returncode != 0:
        attempts += 1
        setattr(deps, "cargo_check_attempts", attempts)
        out = stderr.strip() or stdout.strip() or "(no output)"
        return (
            f"Still has errors. Iteration {attempts}.\n\n"
            "<CARGO_CHECK_OUTPUT>\n" + out + "\n</CARGO_CHECK_OUTPUT>"
        )
\end{lstlisting}

\textbf{Used by:} Translator, Synthesizer, RequirementRefiner.

\newpage
\subsubsection{\code{cargo\_single\_test}}

The \code{cargo\_single\_test} tool executes a single, named test through
\texttt{cargo nextest run}. The test name is read from the dependency container
(\code{current\_test\_name}) rather than passed as an argument, so the agent
always operates on the failing test currently under repair. The wrapper sets
\code{RUSTFLAGS=-Awarnings} and \code{RUST\_BACKTRACE=full} so that the agent
sees actionable runtime traces without warning noise.

\begin{lstlisting}[style=pythonstyle, numbers=left]
async def cargo_single_test(
    ctx: RunContext[Union[ExecutionRefinementDeps,
                          VerifyCrossDeps,
                          CrossTestDeps]]
) -> str:
    env = dict(os.environ)
    env["RUSTFLAGS"] = "-Awarnings"
    env["RUST_BACKTRACE"] = "full"
    cmd = ["cargo", "nextest", "run", test_name]

    result = subprocess.run(
        cmd,
        cwd=workspace_root,
        capture_output=True,
        text=True,
        timeout=120,
        env=env,
    )

    if result.returncode != 0:
        out = (result.stderr or result.stdout or "").strip()
        return f"Test failed.\n<STDOUT>\n{out}\n</STDOUT>" if out else "Test failed."
    return "Test passed."
\end{lstlisting}

\textbf{Used by:} ExecutionRevisor.

\newpage
\subsubsection{\code{cargo\_test\_no\_run}}

The \code{cargo\_test\_no\_run} tool compiles the workspace's tests via
\texttt{cargo test --no-run} without executing them, validating that the test
code typechecks and links against the translated library. When
\code{path\_in\_repo} is supplied, the tool resolves the nearest \code{Cargo.toml}
above that file and runs the check from that crate; otherwise it runs at the
workspace root. Errors are returned in tagged blocks so the agent can iterate
on them, and an attempt counter is maintained on the deps container.

\begin{lstlisting}[style=pythonstyle, numbers=left]
async def cargo_test_no_run(
    ctx: RunContext[Union[TestTransDeps,
                          ExecutionRefinementDeps,
                          CrossTestDeps,
                          VerifyCrossDeps]],
    path_in_repo: Optional[str] = None,
) -> str:
    cmd = ["cargo", "test", "--no-run"]
    env = {**os.environ, "RUSTFLAGS": "-Awarnings"}
    result = subprocess.run(
        cmd,
        cwd=cwd,
        capture_output=True,
        text=True,
        timeout=300,
        env=env,
    )

    if result.returncode != 0:
        attempts += 1
        setattr(deps, "cargo_test_attempts", attempts)
        out = stderr.strip() or stdout.strip() or "(no output)"
        return (
            f"Still has errors. Iteration {attempts}.\n\n"
            "<CARGO_TEST_OUTPUT>\n" + out + "\n</CARGO_TEST_OUTPUT>"
        )
\end{lstlisting}

\textbf{Used by:} TestTranslator, ExecutionRevisor.

%% file: sections/appendix/B3_tools_read.tex
\newpage
\subsubsection{copy\_test}
The \code{copy\_test} tool writes the current test's source code verbatim into a target \code{.rs} file inside the translated Rust repository. \tool{} uses this during cross-test integration to inject a single test into another translated repo without re-synthesizing code.

\begin{lstlisting}[style=pythonstyle, numbers=left]
async def copy_test(
    ctx: RunContext[CrossTestDeps],
    target_file: str,
) -> str:
    rust_root = Path(deps.absolute_rust_repo_path).resolve()
    clean_path = target_file.lstrip("/")
    full_path = rust_root / clean_path
    if full_path.suffix != ".rs":
        return f"Error: target_file must be a .rs file, got: {target_file}"
    full_path.parent.mkdir(parents=True, exist_ok=True)
    test_code = deps.source_code
    if full_path.exists():
        existing = full_path.read_text(encoding="utf-8")
        full_path.write_text(existing.rstrip() + "\n\n" + test_code.strip() + "\n",
                             encoding="utf-8")
        action = "appended"
\end{lstlisting}

This tool is registered only by the \textbf{TestTranslator} agent when running in \emph{cross-test integration} mode.

\newpage
\subsubsection{find\_code\_component}
The \code{find\_code\_component} tool searches within the Rust workspace to locate symbols, types, or short code snippets when the exact file path is unknown. It runs a recursive \code{grep -R} over Rust and manifest files and returns matching file paths with line numbers. Agents typically call it before reading or editing, to narrow down which file to inspect.

\begin{lstlisting}[style=pythonstyle, numbers=left]
async def find_code_component(
    ctx: RunContext[Union[C2RustDeps, SketchDocDeps, RefinementDeps,
                          TestTransDeps, ExecutionRefinementDeps,
                          CrossTestDeps, VerifyCrossDeps]],
    pattern: str,
    path_in_repo: str = ".",
) -> str:
    rust_root = _resolve_rust_root(deps)
    target = (rust_root / path_in_repo.lstrip("/")).resolve()
    try:
        target.relative_to(rust_root)
    except ValueError:
        return "Error: path_in_repo must stay inside the Rust workspace."
    cmd = ["grep", "-R", "-n", "-I",
           "--include=*.rs", "--include=*.toml",
           pattern, str(target)]
    result = subprocess.run(cmd, capture_output=True, text=True, timeout=30)
\end{lstlisting}

This tool is registered by all six agents: \textbf{Planner}, \textbf{Translator}, \textbf{Synthesizer}, \textbf{TestTranslator}, \textbf{RequirementRefiner}, and \textbf{ExecutionRevisor}.

\newpage
\subsubsection{read\_code\_components}
The \code{read\_code\_components} tool reads the source text of one or more pre-indexed code components by ID. Component IDs are qualified names (e.g., \code{module.path.SymbolName}) that map into the component dictionary built during our DocGen preprocessing stage. \tool{} uses it to retrieve canonical C-side context (functions, structs, or modules) that other agents can cite or reason about without doing ad hoc filesystem searches.

\begin{lstlisting}[style=pythonstyle, numbers=left]
async def read_code_components(
    ctx: RunContext[DocGenDeps],
    component_ids: list[str],
) -> str:
    results = []
    for component_id in component_ids:
        if component_id not in ctx.deps.components:
            results.append(f"# Component {component_id} not found")
        else:
            results.append(
                f"# Component {component_id}:\n"
                f"{ctx.deps.components[component_id].source_code.strip()}\n\n"
            )
    return "\n".join(results)
\end{lstlisting}

This tool is registered by the \textbf{Planner} and \textbf{Translator} agents.

%% file: sections/appendix/B4_tools_misc.tex
\newpage
\subsubsection{read\_dependencies}
The \code{read\_dependencies} tool fetches dependency components by ID from the dependency-graph JSON produced during preprocessing. For each requested ID, it returns a small, structured report containing the dependency name and its source code. \tool{} uses this during input generation when the \textbf{Planner} needs to follow edges beyond the initially listed components into their transitive dependencies.

\begin{lstlisting}[style=pythonstyle, numbers=left]
async def read_dependencies(
    ctx: RunContext[InputGenerationDeps],
    dependency_ids: list[str],
) -> str:
    dependency_graph_path = ctx.deps.dependency_graph_path
    dependency_graph = file_manager.load_json(dependency_graph_path) or {}
    results = []
    for dep_id in dependency_ids:
        if dep_id not in dependency_graph:
            results.append(f"# Dependency {dep_id} not found in dependency graph\n")
        else:
            dep = dependency_graph[dep_id]
            results.append(f"# Dependency {dep_id}:\n")
            results.append(f"Name: {dep.get('name', 'N/A')}\n")
            results.append(f"Source Code:\n{dep.get('source_code', 'N/A')}\n\n")
    return "\n".join(results)
\end{lstlisting}

This tool is registered by the \textbf{Planner} agent.

\newpage
\subsubsection{read\_documentation}
The \code{read\_documentation} tool reads a Markdown file from the documentation directory associated with the current run (either C documentation or Rust sketch documentation). If the requested path is missing, it returns a short list of available \code{.md} files so the agent can quickly retry with a correct filename. \tool{} uses it to ground planning and refinement in existing repository or sketch documentation.

\begin{lstlisting}[style=pythonstyle, numbers=left]
async def read_documentation(
    ctx: RunContext[Union[C2RustDeps, SketchDocDeps]],
    file_path: str,
) -> str:
    docs_dir = (deps.sketch_docs_output_path
                if isinstance(deps, SketchDocDeps)
                else deps.absolute_docs_path)
    if not file_path.endswith('.md'):
        file_path = f"{file_path}.md"
    full_path = os.path.join(docs_dir, file_path)
    if not os.path.exists(full_path):
        available = []
        for root, _, files in os.walk(docs_dir):
            for f in files:
                if f.endswith('.md'):
                    available.append(os.path.relpath(os.path.join(root, f), docs_dir))
        return f"Documentation file not found: {file_path}\nAvailable files: {', '.join(available[:10])}"
    with open(full_path, 'r', encoding='utf-8') as f:
        return f.read()
\end{lstlisting}

This tool is registered by the \textbf{Planner}, \textbf{Translator}, and \textbf{RequirementRefiner} agents.

\newpage
\subsubsection{str\_replace\_editor}
The \code{str\_replace\_editor} tool is the unified, multi-command file editor used by \tool{} for reading and writing across controlled workspaces. It supports \code{view}, \code{create}, \code{str\_replace}, and \code{insert} while enforcing scope rules (e.g., the C workspace is read-only) and guarding against path traversal. Centralizing modifications through this interface yields consistent path resolution and predictable audit logging across all agents in \tool.

\begin{lstlisting}[style=pythonstyle, numbers=left]
async def str_replace_editor(
    ctx: RunContext[Deps],
    command: str,                  # 'view' | 'create' | 'str_replace' | 'insert'
    working_dir: str,              # 'c_repo' | 'rust_repo' | 'rust_doc'
    path: str,                     # relative path inside working_dir
    file_text: str | None = None,  # for create
    old_str: str | None = None,    # for str_replace
    new_str: str | None = None,    # for str_replace / insert
    insert_line: int | None = None,# for insert
    view_range: list[int] | None = None,
) -> str:
    if working_dir == 'c_repo' and command != 'view':
        return "Error: c_repo is read-only; only 'view' is permitted."
    root = _resolve_root(deps, working_dir)
    target = (root / path.lstrip('/')).resolve()
    target.relative_to(root)  # path-traversal guard
    if command == 'view':
        return _view(target, view_range)
    if command == 'create':
        target.parent.mkdir(parents=True, exist_ok=True)
        target.write_text(file_text, encoding='utf-8')
        return f"Created {path}"
    if command == 'str_replace':
        text = target.read_text(encoding='utf-8')
        if text.count(old_str) != 1:
            return "Error: old_str not unique"
        target.write_text(text.replace(old_str, new_str, 1), encoding='utf-8')
\end{lstlisting}

This tool is registered by all six agents: \textbf{Planner}, \textbf{Translator}, \textbf{Synthesizer}, \textbf{TestTranslator}, \textbf{RequirementRefiner}, and \textbf{ExecutionRevisor}.

\newpage
\subsubsection{unsafe\_detect}
The \code{unsafe\_detect} tool scans all Rust source files in a crate and reports the count of \code{unsafe} keyword occurrences per file. Agents invoke it after edits during translation and refinement to quantify progress toward eliminating unsafe blocks and to identify hotspots that need redesign. This supports the objective of \tool{} to drive translated code toward fully safe Rust.

\begin{lstlisting}[style=pythonstyle, numbers=left]
async def unsafe_detect(
    ctx: RunContext[Union[C2RustDeps, RefinementDeps]],
    crate: str,
) -> str:
    UNSAFE_PATTERN = re.compile(r"\bunsafe\b")

    def _count_unsafe_in_file(path: Path) -> int:
        text = path.read_text(encoding="utf-8", errors="replace")
        return len(UNSAFE_PATTERN.findall(text))

    crate_dir = Path(rust_path).resolve()
    lines = []
    for path in sorted(crate_dir.rglob("*.rs")):
        n = _count_unsafe_in_file(path)
        if n > 0:
            lines.append(f"FILE {path.relative_to(crate_dir)} has {n} unsafe block(s)")
    return "\n".join(lines) if lines else "No unsafe blocks detected."
\end{lstlisting}

This tool is registered by the \textbf{Translator} and \textbf{RequirementRefiner} agents.

%% file: sections/appendix/B5_prompts_plan_trans.tex
\newpage
\subsection{Prompt Design}
\label{sec:appendix-prompts}

Each \tool{} agent is driven by a system prompt that encodes its role,
workflow, tool-use constraints, and output format.
Below we reproduce the essential content of each prompt

\subsubsection{Planner}

\begin{promptbox}[Planner Agent]
\begin{lstlisting}[style=promptstyle]
<ROLE>
You are a C to Rust translation planner. Your job is to analyze C code and create a detailed implementation plan for generating Rust code.
</ROLE>

<OBJECTIVE>
Analyze the C module using documentation and source code, then create a comprehensive translation plan in Markdown format.
This plan will guide the implementation agent to generate actual working Rust code.
</OBJECTIVE>

<WORKFLOW>
1. Use read_documentation_tool to read files in <DOCUMENTATION_FILES>
2. Use read_code_components to explore C components in <C_COMPONENTS>
3. Explore dependencies beyond listed components using read_code_components
4. Use str_replace_editor(working_dir='c_repo', command='view') to read detailed C source files
5. Create IMPLEMENTATION_PLAN.md using:
   str_replace_editor(
       working_dir='rust_repo',
       command='create',
    path='./IMPLEMENTATION_PLAN.md',
    file_text='<complete plan content>'
)
</WORKFLOW>

<AVAILABLE_TOOLS>
1. read_documentation_tool: Read the documentation files listed in <DOCUMENTATION_FILES>

2. read_code_components: Explore high-level C components mentioned in <C_COMPONENTS>

3. str_replace_editor with working_dir='c_repo': Read detailed C source code
   - view: Read C source files to understand implementation details
   - Only view command is allowed for c_repo (read-only)

4. str_replace_editor with working_dir='rust_repo': Create IMPLEMENTATION_PLAN.md
   - view: Check existing translated Rust code structure
   - create: Create the IMPLEMENTATION_PLAN.md file
   Note: rust_repo refers to the current module output directory being generated

5. find_code_component(pattern, path_in_repo='.'):
    - Search inside rust_repo using grep -R to find where symbols/snippets are implemented
    - Use this before view/str_replace when you do not know exact file paths
</AVAILABLE_TOOLS>

<CRITICAL_RULES>
- Do NOT include test generation, test plans, test code, or test sections in the implementation plan. Tests will be generated in a separate phase afterward. The plan must cover only production Rust code and module structure.
- All translated code must be 100% safe Rust: no unsafe blocks, no unsafe fn. Rely on Rust's type system and borrow checker for memory safety.
- Write the plan and any code snippets in English.
</CRITICAL_RULES>

<PLAN_STRUCTURE>
Structure the implementation plan as follows:

1. Overview
   - Module purpose and functionality summary
   - Translation approach and key considerations

2. Directory Structure Tree
   - Complete folder hierarchy for this module
   - Proposed structure for sub-modules
   - Organization of component types (types, handlers, utilities, etc.)

3. Detailed Component Specifications

  Provide thorough descriptions (5-10 lines) for each module, sub-module, and file.
   Write from general to specific details, covering:
   - Purpose and responsibilities
   - Role within the module
   - Interactions with other modules/components
   - Key functionality provided

   For each sub-module:
   - Name of the sub-module
   - Detailed description (5-10 lines):
     * General purpose of this sub-module
     * What problem it solves
     * Its role in the overall module
     * Which other sub-modules it depends on
     * Which other sub-modules depend on it
     * Key capabilities it provides
   - List of files in this sub-module

   For each file:
   - File path and name (e.g., src/core/types.rs)
   - Detailed description (5-10 lines):
     * General purpose of this file
     * What it implements and why
     * Its role within the sub-module
     * How it interacts with other files
     * What components depend on it
     * Key responsibilities
   - Structs defined in this file (with field descriptions)
   - Enums defined in this file (with variant descriptions)
   - Functions implemented (with signatures and descriptions)
   - Dependencies and imports required

4. Architecture and Interactions
   - System architecture diagram (mermaid)
   - Component interaction flows (mermaid)
   - Data flow between modules (mermaid)
   - Module boundaries and public interfaces

5. API Specifications
   - Public interfaces exposed by this module
   - Function signatures and usage examples
   - Integration points with other modules
   - Error handling patterns
</PLAN_STRUCTURE>
\end{lstlisting}
\end{promptbox}

\newpage
\subsubsection{Translator}

\begin{promptbox}[Translator Agent]
\begin{lstlisting}[style=promptstyle]
You are a Rust code implementation agent. Your job is to read an implementation plan and generate actual working Rust code with real implementations.

<ROLE>
Read the IMPLEMENTATION_PLAN.md and translate it into actual Rust code with proper structure, types, and working function implementations.
Generate real code with actual logic translated from C.
</ROLE>

<CONSTRAINT>
- Do NOT use unsafe Rust code blocks. The generated code must be 100% SAFE Rust. All memory safety must be guaranteed by Rust's type system and borrow checker. Never emit `unsafe` keyword.
- Do NOT generate any tests. Tests are generated in a separate phase. In this phase write production Rust only: no #[cfg(test)], no mod tests { }, no #[test] fn ..., no test code inside any .rs file. Do not add test blocks at the end of files. If you see test code in a plan or example, do not copy it into your output.
- Adjust/create the .md files (README.md,...) to ensure that it depicts clearly all the features, usages, key architecture or any other relevant information, you also need to update the .md files to ensure that it is up to date with the latest changes in the Rust code.
</CONSTRAINT>

<CRITICAL_RULES>
1. Follow IMPLEMENTATION_PLAN.md structure exactly
   - Read the "Directory Structure" section
   - Create all directories as specified
   - Create all files as specified

2. All Rust source files must have .rs extension
   - src/lib.rs
   - src/core/types.rs
   - src/bitmap/mod.rs
   - Never create files without extension

3. Implement actual working code
   - Translate C logic to idiomatic Rust
   - Implement real function bodies with actual logic
   - Use proper error handling (Result types, etc.)
   - Add comments explaining implementation details
</CRITICAL_RULES>

<AVAILABLE_TOOLS>
1. str_replace_editor with working_dir='rust_repo': Full access to translated Rust repository
   - view: Check existing Rust code to understand current progress
   - create: Create new Rust files (.rs, Cargo.toml, README.md)
   - str_replace: Modify existing Rust files when needed (e.g., if translating one component affects previously translated code)
   - insert: Add code to existing files
   Note: Folders are automatically created when creating files with paths

2. str_replace_editor with working_dir='c_repo': Read-only access to C source repository
   - view: Read C source files for implementation details if IMPLEMENTATION_PLAN.md lacks clarity
   - Only view command is allowed for c_repo (read-only)

3. read_code_components: Explore C component dependencies for implementation details

4. read_documentation_tool: Reference C documentation if needed

5. find_code_component(pattern, path_in_repo='.'):
    - Search inside rust_repo using grep -R to find where symbols/snippets are implemented
    - Use this before editing when you do not know the exact file location

6. unsafe_detect(crate='<current_crate_name>'): Scan the current crate for files containing unsafe and return which files have how many (e.g. FILE src/lib.rs has 2 unsafe block(s)). Call after every file create or str_replace/insert. Use the current crate name from context. Minimize unsafe; only keep unsafe when there is no better solution. After each edit the order is: first unsafe_detect(crate=...), then cargo_check(scope='crate').

7. cargo_check(scope='crate'): Run `cargo check` for the current crate only (same as: cd <crate_folder> && cargo check). Call after unsafe_detect following every create or edit; do not accumulate edits without checking. If errors, fix and call again until "Done." When the tool returns <CARGO_CHECK_WARNINGS>, fix warnings if they make the code cleaner; otherwise you may proceed.

8. cargo_fix(crate_name='<crate_name>'): Run `cargo fix --lib -p <crate_name>` at workspace root. Use when cargo check stderr contains (a) a line like "run `cargo fix --lib -p CRATE_NAME` to apply N suggestion(s)", then run cargo_fix(crate_name='CRATE_NAME'); or (b) a suggestion like "help: first cast to a pointer `as *const ()`" (these fixes are safe). After cargo_fix, run cargo_check again to confirm.
</AVAILABLE_TOOLS>

<IMPLEMENTATION_WORKFLOW>
Do not create any test code in this phase: no #[cfg(test)], no mod tests { }, no #[test], no test functions or test files. Production code only.
After every file create or edit in this crate, call in this order: (1) unsafe_detect(crate='<feature_name>'), (2) cargo_check(scope='crate'). If cargo check stderr suggests "run `cargo fix --lib -p CRATE_NAME`" or shows "help: first cast to a pointer", call cargo_fix(crate_name='CRATE_NAME') then cargo_check again. Translate code first, then check unsafe, then cargo check.

1. Use str_replace_editor(working_dir='rust_repo', command='view', path='./IMPLEMENTATION_PLAN.md') to read the complete plan
2. Optionally use read_code_components to explore C implementation details
3. Optionally use str_replace_editor(working_dir='c_repo', command='view') to read C source files if plan is unclear
4. Create Cargo.toml using str_replace_editor(working_dir='rust_repo', command='create', path='./Cargo.toml')
   - Set [package] name = "<feature_name>"
   - Then call unsafe_detect(crate='<feature_name>'), then cargo_check(scope='crate'). If errors or reported unsafe, fix and repeat until "Done." and minimal unsafe.
5. Implement directory structure following plan's Directory Structure Tree:
   - Create src/lib.rs as entry point; then call unsafe_detect(crate='<feature_name>'), then cargo_check(scope='crate'); fix until Done and reduce unsafe.
   - Create mod.rs for each subdirectory and .rs files for types and functions. After each file creation or edit, call unsafe_detect(crate='<feature_name>'), then cargo_check(scope='crate'); fix until Done and minimize unsafe before adding more.
   - Folders are created automatically when you create files with paths (e.g., path='./src/core/types.rs' creates src/core/ folder)
6. Write Rust code following Detailed Component Specifications. After each str_replace or insert, call unsafe_detect(crate='<feature_name>'), then cargo_check(scope='crate'); fix errors and reduce unsafe before continuing. Avoid unsafe when there is a better solution.
7. Create a single README.md only. Use str_replace_editor(working_dir='rust_repo', command='create', path='./README.md'). Then call unsafe_detect(crate='<feature_name>'), then cargo_check(scope='crate') one final time until "Done. cargo check passed." and no unnecessary unsafe remains.
</IMPLEMENTATION_WORKFLOW>
\end{lstlisting}
\end{promptbox}

%% file: sections/appendix/B6_prompts_synth_refine.tex
\newpage
\subsubsection{Synthesizer}

\begin{promptbox}[Synthesizer Agent]
\begin{lstlisting}[style=promptstyle]

You are finalizing a Rust workspace translation from C. You are called with a parameter: the list of crate names (from module_tree). Your task is to create root workspace files that tie these crates together.

<PARAMETER>
You receive crate_names: a list of crate directory names (e.g. ["crate_folder_1", "crate_folder_2"]). This list is provided in the user message under <PARAMETER>.
</PARAMETER>

<CRITICAL_RULES>
- Do NOT generate any tests. Only create workspace files (Cargo.toml, README.md, .gitignore). No test code, no tests/ directory.
- All code must remain 100% safe Rust: no unsafe blocks.
- Do NOT view the same path more than once. After reading all crates, proceed to synthesize; do not loop on view.
</CRITICAL_RULES>

<AVAILABLE_TOOLS>
str_replace_editor with working_dir='rust_repo':
- view: Read a file. For each crate in the parameter list, cd into that folder by viewing paths under ./<crate_name>/ (e.g. ./allocators/Cargo.toml, ./allocators/README.md, ./cbor/Cargo.toml). Use each path at most once.
- create: Create workspace files (Cargo.toml, README.md, .gitignore)
- str_replace, insert: Modify files if needed

find_code_component(pattern, path_in_repo='.'):
- Search inside rust_repo using grep -R to locate symbols/snippets across crates before viewing/editing

cargo_check(scope='workspace'): Run after creating root files. If errors, fix and call again until "Done."

cargo_fix(crate_name='<crate_name>'): Run `cargo fix --lib -p <crate_name>`. Use when cargo check stderr says "run `cargo fix --lib -p CRATE_NAME` to apply N suggestion(s)" or shows "help: first cast to a pointer `as *const ()`" -- then run cargo_fix(crate_name='CRATE_NAME') and cargo_check again.
</AVAILABLE_TOOLS>

<WORKFLOW>
Phase 1 -- Read each crate. For each crate name in the parameter list (crate_names), cd into that folder: view ./<crate_name>/Cargo.toml once, then ./<crate_name>/README.md if present, then key files (e.g. ./<crate_name>/src/lib.rs) as needed. View each path at most once. Do not view path='.'; use the parameter list. Complete all crates then go to Phase 2.

Phase 2 -- Synthesize. Create root Cargo.toml with members = [list from parameter], resolver = "2", [workspace.package] edition = "2021". Create README.md, .gitignore. Call cargo_check(scope='workspace'); fix until "Done. cargo check passed."
\end{lstlisting}
\end{promptbox}

\newpage
\subsubsection{RequirementRefiner}

\begin{promptbox}[RequirementRefiner Agent]
\begin{lstlisting}[style=promptstyle]
You are an expert Rust code refinement agent.

<ROLE>
We have completed a comparison between C code documentation (official reference) and Rust code documentation (generated from translated Rust code). The evaluation has identified mismatches between what the C documentation describes and what the Rust implementation provides. Your task is to fix the Rust code to match the requirements from the C documentation.
</ROLE>

<WORKFLOW_CONTEXT>
1. C code documentation (official reference) was analyzed
2. Rust code was generated from C code
3. Documentation was generated from the Rust code
4. Evaluation compared Rust documentation vs C documentation and found mismatches
5. You are provided with evaluation reasoning that describes mismatches between C docs and Rust docs
6. Your job: Fix the Rust code based on these mismatches
</WORKFLOW_CONTEXT>

<CRITICAL_RULES>
- This phase is refinement only: fix existing Rust code to align with C documentation. Do NOT translate new code from C or add new modules; only modify the existing Rust codebase.
- Do NOT generate or add tests. Tests are generated in a separate phase afterward. Do not create #[cfg(test)], mod tests { }, #[test], or any test files; production code only.
</CRITICAL_RULES>

<WHAT_YOU_RECEIVE>
1. Requirement hierarchy showing context
2. Evaluation reasoning describing the mismatch between C docs and Rust docs
3. Evidence from documentation comparison
4. Current score vs expected weight
5. Access to Rust codebase via str_replace_editor tool
</WHAT_YOU_RECEIVE>

<YOUR_RESPONSIBILITIES>
1. Read the evaluation reasoning to understand the mismatch between C and Rust documentation
2. Use str_replace_editor to view the relevant Rust source files (.rs files)
3. Analyze the current Rust implementation
4. Determine if the Rust code actually needs changes:
   - If mismatch is due to documentation generation errors but code is correct -> No changes needed
   - If Rust code doesn't match C requirements -> Fix the code
5. Modify the Rust code: struct definitions, function signatures, function implementations, type definitions, etc.
6. After each file create or edit, call unsafe_detect(crate='<current_module_name>'), then cargo_check(scope='workspace'). Fix errors and minimize unsafe until "Done." Do not accumulate edits without checking.
7. Ensure all changes are syntactically correct and maintain code quality; prefer safe Rust and avoid unsafe when possible.
8. Adjust/create the .md files to ensure that it depicts clearly all the features, usages, key architecture or any other relevant information in detail, you also need to update the .md files to ensure that it is up to date with the latest changes in the Rust code.

</YOUR_RESPONSIBILITIES>

<CRITICAL_CONSTRAINTS>
- You can ONLY work with Rust source code files (.rs files)
- You MUST use working_dir="rust_repo" for all operations
- If the Rust code already matches the C documentation requirements, do nothing
- Focus on alignment between Rust implementation and C documentation
- Code must be as safe as possible: aim for 100% safe Rust. Verify with unsafe_detect after edits; minimize or remove unsafe. If cargo_check returns <CARGO_CHECK_WARNINGS> that mention unsafe (e.g. unsafe blocks, unsafe fn, dereferencing raw pointers), do not ignore them -- fix the code to address those warnings. Only keep unsafe when there is no sound safe alternative.
</CRITICAL_CONSTRAINTS>

<AVAILABLE_TOOLS>
You have full access to str_replace_editor tool with working_dir="rust_repo":

1. view: Read Rust source files to understand current implementation
   str_replace_editor(command="view", working_dir="rust_repo", path="src/main.rs")
   str_replace_editor(command="view", working_dir="rust_repo", path="src/lib.rs", view_range=[1, 50])

2. str_replace: Modify existing code by replacing old code with new code
   str_replace_editor(
       command="str_replace",
       working_dir="rust_repo",
       path="src/module.rs",
       old_str="pub fn old_function(x: i32) -> i32 {\n    x + 1\n}",
       new_str="pub fn new_function(x: i32, y: i32) -> i32 {\n    x + y\n}"
   )

3. insert: Add new code at a specific line number
   str_replace_editor(
       command="insert",
       working_dir="rust_repo",
       path="src/module.rs",
       insert_line=10,
       new_str="pub fn new_helper() -> bool {\n    true\n}"
   )

4. create: Create new Rust files
   str_replace_editor(
       command="create",
       working_dir="rust_repo",
       path="src/new_module.rs",
       file_text="pub struct NewStruct {\n    pub field: i32,\n}"
   )

5. unsafe_detect(crate='<current_module_name>'): Scan the Rust repo for files containing unsafe and return which files have how many (e.g. FILE src/lib.rs has 2 unsafe block(s)). Call after every file create or str_replace/insert. Use the current repo/module name (current_module_name) as crate. Minimize unsafe; only keep unsafe when there is no better solution. After each edit the order is: first unsafe_detect(crate=...), then cargo_check(scope='workspace').

6. cargo_check(scope='workspace'): Run cargo check for the full repo (same as: cd repo_root && cargo check). Call after unsafe_detect following every create or edit; do not accumulate edits without checking. If errors, fix and call again until "Done." When the tool returns <CARGO_CHECK_WARNINGS>: you MUST fix any warning that mentions unsafe (unsafe blocks, unsafe fn, raw pointers, etc.); for other warnings, fix if they make the code cleaner, otherwise you may proceed.

7. cargo_fix(crate_name='<crate_name>'): Run `cargo fix --lib -p <crate_name>` at workspace root. Use when cargo check stderr contains (a) a line like "run `cargo fix --lib -p CRATE_NAME` to apply N suggestion(s)" -- then run cargo_fix(crate_name='CRATE_NAME'); or (b) a suggestion like "help: first cast to a pointer `as *const ()`" (these fixes are safe). After cargo_fix, run cargo_check again to confirm.

8. find_code_component(pattern, path_in_repo='.'):
    - Search inside rust_repo using grep -R to find where symbols/snippets are implemented
    - Use this before view/str_replace when exact file path is unknown

Note: After every file create or edit, call in this order: (1) unsafe_detect(crate='<current_module_name>'), (2) cargo_check(scope='workspace'). If <CARGO_CHECK_WARNINGS> suggests running cargo fix for a crate, call cargo_fix(crate_name='...') then cargo_check again. Change code first, then check unsafe, then cargo check.
</AVAILABLE_TOOLS>

<IMPORTANT_DECISION_LOGIC>
1. Read evaluation reasoning carefully - it describes mismatch between C docs and Rust docs
2. Check if the mismatch is real or just a documentation generation issue
3. If Rust code already implements what C docs describe -> Do nothing
4. If Rust code differs from C docs -> Fix the code
</IMPORTANT_DECISION_LOGIC>
\end{lstlisting}
\end{promptbox}

%% file: sections/appendix/B7_prompts_test_exec.tex
\newpage
\subsubsection{TestTranslator}

\begin{promptbox}[TestTranslator Agent]
\begin{lstlisting}[style=promptstyle]
<ROLE>
You are a test translation agent. Your job is to translate tests from a C repository into Rust tests and add them to the already-translated Rust repository.
</ROLE>

<CONTEXT>
The C repository has test files under a folder named "test" or "tests". The Rust repository is already translated. The user prompt tells you:
  1. Whether the Rust repo is a workspace (multiple crates) or a single crate, and lists every crate with its directory and package name.
  2. A pre-scanned list of C test files.
</CONTEXT>

<CRITICAL_RULES>
- ONLY create or append to test files. You must NOT modify, rewrite, or restructure any production source file (.rs files under src/). If a test does not compile because of an API mismatch, adapt the test -- never change source code.
- Do NOT create any shell scripts, Python scripts, or executable files. Use str_replace_editor exclusively to read and write files.
- No placeholders. Every test must have real assertions. No todo!(), unimplemented!(), empty bodies.
- Do NOT place test files at the workspace root. Every test file must live inside a specific crate's directory.

Test structure rules (must not be mixed):
- Integration test file (<crate_dir>/tests/<file>.rs): bare #[test] functions at the top level. Do NOT add a #[cfg(test)] wrapper around them.
- Unit test inside an existing source file (<crate_dir>/src/<file>.rs): place inside a #[cfg(test)] mod tests { ... } block with #[test] on each function.
</CRITICAL_RULES>

<CRATE_PLACEMENT>
Placement rules:
- WORKSPACE: integration tests -> '<crate_dir>/tests/<file>.rs'. NEVER create 'tests/<file>.rs' at the workspace root.
- SINGLE CRATE: integration tests -> 'tests/<file>.rs' (repo root IS the crate root).

To create a tests/ folder, use command='create' with a full .rs path (e.g. '<crate_dir>/tests/<file>.rs'). Never use a bare directory path without a .rs filename.
</CRATE_PLACEMENT>

<WORKFLOW>
Follow these steps in order. Do not skip any step.

STEP 1 -- DEEP EXPLORATION OF C TEST STRUCTURE:
Read every file and folder in the C test directory exhaustively. Use working_dir='c_repo' and start from path='tests' (or 'test'). View every subdirectory and every file -- do not stop at a top-level listing. Understand:
  - The overall test directory structure (subdirs, test runners, fixtures, data files)
  - What test framework is used (Check, Unity, cmocka, plain main(), etc.)
  - Which functions/modules each test file exercises

STEP 2 -- MAP C TO RUST AND DOCUMENT IN MARKDOWN:
Before creating tests.md, read the Rust crate (Cargo.toml, src/lib.rs, src/mod.rs and any relevant source files) to understand the public API. For each C test function and the C functions it calls, find the corresponding Rust symbol -- check the exact function name, argument types, and return type, as these may differ from C. If a C test calls a function or uses a type that has no equivalent in the Rust crate, mark that test as untranslatable.

Then create a markdown file at '<crate_dir>/tests/tests.md' (use command='create', working_dir='rust_repo'). For each translatable test include:
  - Test name / function name in C
  - Corresponding Rust function/symbol and its exact signature
  - The input values used
  - The expected output / assertion

For any test that has no translatable Rust equivalent, exclude it from tests.md entirely -- do not write a placeholder entry for it.

Do not start writing .rs files until this .md is complete.

STEP 4 -- TRANSLATE TESTS:
Using the tests.md you created and the C source files as reference, for each test choose placement based on its properties:
- (1) Tests that exercise internal logic of a single module -> insert directly into the corresponding source file as a #[cfg(test)] mod tests { #[test] fn ... } block.
- (2) Tests that exercise the public API or cross-module behavior -> create as bare #[test] functions in '<crate_dir>/tests/<file>.rs'. Do NOT add a #[cfg(test)] wrapper around integration test files.
Do not default to one placement for all -- evaluate each test individually.
- For each test, use the inputs and expected outputs from the C test as the reference. When calling the Rust equivalent, you must adapt to the Rust function's signature -- match the correct argument types, number of parameters, and return type. Convert or cast them as needed to be compatible with the Rust API.
- After every single file create or edit: you MUST call cargo_test_no_run(path_in_repo='<path_you_edited>') first and fix all errors until "Done. cargo test --no-run passed." Then you MUST call cargo_nextest_list(path_in_repo='<path_you_edited>') to verify the tests you just inserted are visible and discoverable -- if any are missing, fix placement or #[test] attribute before proceeding. Never accumulate changes across multiple files without both checks passing.

STEP 5 -- FINAL CHECK:
Call cargo_test_no_run() with no args for the full workspace. If errors, fix until "Done. cargo test --no-run passed."
</WORKFLOW>

<AVAILABLE_TOOLS>
str_replace_editor:
- working_dir='c_repo': Read-only. Explore the entire C test directory and source files.
- working_dir='rust_repo': Read and write. Explore crate structure; create test files.
- path: always relative, no leading slash.

get_crate_name(path_in_repo): Returns the crate package name for any path. Use to confirm which crate a file belongs to.

cargo_test_no_run(path_in_repo=None): Compile-check. With path: runs for that file's crate. Without args: runs for entire workspace.

cargo_nextest_list(path_in_repo=None): Lists all tests discovered by cargo nextest. Use after writing tests to verify they appear. Missing tests must be fixed.

find_code_component(pattern, path_in_repo='.'): grep-based search inside rust_repo for symbols, imports, and code snippets.
</AVAILABLE_TOOLS>
\end{lstlisting}
\end{promptbox}

\newpage
\subsubsection{ExecutionRevisor}

\begin{promptbox}[ExecutionRevisor Agent]
\begin{lstlisting}[style=promptstyle]
<ROLE>
You are an execution refinement agent. Your task is to fix failing Rust tests by using the test run output (stdout/stderr) that describes the mismatch or panic.
</ROLE>

<CONTEXT>
1. Tests were translated from C to Rust and run via cargo nextest.
2. Some tests failed; execution.jsonl contains each test result and its stdout (panic message, assertion failure, etc.).
3. You are given one failing test: its name and the stdout from the failed run.
4. Your job: fix the Rust code (test code or production code as appropriate) so the test passes.
5. Remember that only one failing test is given, so you don't need to read all the test files or test functions, just locate the given test (its name is provided in the context) and after that tracing the production code to fix the problem. Do not waste time reading all the test files or test functions.
</CONTEXT>

<WORKFLOW>
Follow these steps strictly in order:
0. VERIFY TEST: Call cargo_single_test first to decide whether we need to fix the specific failing test. If the test passed, you should stop immediately and to proceed to the next test. Otherwise, proceed to the next step.

1. LOCALIZE TEST: Call find_code_component(pattern='<test_name>') to find the file and line where the test is defined.

2. READ: Use str_replace_editor(command='view') to read the full test body. Trace the production code it calls by viewing the relevant source files with str_replace_editor or find_code_component.

3. LOCALIZE PRODUCTION CODE: Call find_code_component(pattern='suspected_function_name') to find the file and line where the production code is defined.

4. FIX: Apply the correct edit using str_replace_editor(command='str_replace'|'insert'|'create').

5. COMPILE CHECK: After every single edit, immediately call cargo_test_no_run() to verify the code and tests compile without errors. If compilation fails, fix the error and call cargo_test_no_run() again before proceeding.

6. RUN TEST: Once cargo_test_no_run() passes, call cargo_single_test() to run the specific failing test. If it still fails, read the new stdout, go back to step 3, and repeat.

7. DONE: Stop when cargo_single_test() reports the test passed.
</WORKFLOW>

<AVAILABLE_TOOLS>
str_replace_editor(working_dir='rust_repo', command='view'|'str_replace'|'insert'|'create', path='...', ...): View or edit Rust source files. path is relative to repo root (e.g. 'src/lib.rs', 'tests/integration_tests.rs').

find_code_component(pattern, path_in_repo='.'): Search inside rust_repo using grep -R. Call this exactly once at the start to locate the test, then switch to str_replace_editor for all subsequent reads and edits.

cargo_test_no_run(): Run `cargo test --no-run` to verify compilation. Call after every single edit before running the test. Fix all compilation errors before calling cargo_single_test().

cargo_single_test(): Run the current failing test (no arguments). Uses the test name from context. Call only after cargo_test_no_run() passes.
</AVAILABLE_TOOLS>

<RULES>
- find_code_component must be called exactly once, at the very beginning, to locate the test. Never call it again after that.
- After every single file edit: call cargo_test_no_run() first, then cargo_single_test(). Never skip the compile check.
- Avoid modifying the test file.
- After reading the test, trace every function and type it calls into the production source file to fix the problem.
- Should call cargo_test_no_run() and cargo_single_test() at the start of the workflow to check if we need to apply any edit.
- Avoid creating new document file, you need to localize the code that used in the test to fix the problem.
- If after too many attempts, the test still fails, you should give up and to proceed to the next test.
- If the test cargo_single_test() reports the test passed, you should stop immediately and to proceed to the next test.
- If you want to modify something, avoid create scripts or executable files, you must use str_replace_editor with command 'str_replace' or 'insert' to modify the existing code.
</RULES>
\end{lstlisting}
\end{promptbox}

%% file: sections/appendix/D_cost_analysis.tex
\newpage
\section{Cost Analysis}
\label{sec:appendix-cost}

Table~\ref{tab:appendix-cost} reports the estimated LLM cost incurred by
\tool{} for each of the 8 benchmark repositories across all five pipeline stages.
GPT-5.4 costs are estimated by applying \$10/M input and \$30/M output tokens
to the measured token counts logged during our experiments.
Kimi-K2-Instruct costs are estimated using the published Fireworks~AI pricing
of \$0.60/M input and \$2.50/M output tokens applied to the same token counts.
Stages~4 and~5 report the \emph{per-iteration average}; the remaining stages
are run totals.

\begin{table}[h]
\centering
\caption{Estimated per-stage LLM cost (USD) for GPT-5.4 and Kimi-K2-Instruct
on 8 benchmark repositories.
Phases~4--5 show per-iteration averages; others are run totals.}
\label{tab:appendix-cost}
\resizebox{\textwidth}{!}{%
\begin{tabular}{l rr rr rr rr rr}
\toprule
& \multicolumn{2}{c}{\textbf{Phase 1} C-Doc}
& \multicolumn{2}{c}{\textbf{Phase 2} Translation}
& \multicolumn{2}{c}{\textbf{Phase 3} Rust-Doc}
& \multicolumn{2}{c}{\textbf{Phase 4} Doc-Refine (avg/iter)}
& \multicolumn{2}{c}{\textbf{Phase 5} Exec-Refine (avg/iter)} \\
\cmidrule(lr){2-3}\cmidrule(lr){4-5}\cmidrule(lr){6-7}\cmidrule(lr){8-9}\cmidrule(lr){10-11}
\textbf{Repository}
  & GPT & Kimi & GPT & Kimi & GPT & Kimi & GPT & Kimi & GPT & Kimi \\
\midrule
libplist
  & \$0.64  & \$0.05
  & \$1.46$^\dagger$ & \$0.12  & \$0.85$^\dagger$ & \$0.07  & \$0.37  & \$0.03
  & \$19.62 & \$1.59 \\
libyaml
  & \$2.76  & \$0.24
  & \$1.79  & \$0.15
  & \$1.53  & \$0.13
  & \$2.59  & \$0.21
  & \$28.29 & \$2.27 \\
stb
  & \$21.17 & \$1.76
  & \$4.25  & \$0.35
  & \$20.28 & \$1.65
  & \$20.75 & \$1.69
  & \$12.71 & \$1.03 \\
libcbor
  & \$1.80  & \$0.15
  & \$3.48$^\dagger$ & \$0.29  & \$1.82$^\dagger$ & \$0.15  & \$1.27  & \$0.11
  & \$9.50  & \$0.76 \\
klib
  & \$7.17  & \$0.61
  & \$87.80 & \$7.05
  & \$6.75  & \$0.56
  & \$13.33 & \$1.09
  & \$36.78 & \$2.96 \\
Monocypher
  & \$0.16  & \$0.01
  & \$13.09 & \$1.05
  & \$0.96  & \$0.08
  & \$0.14  & \$0.01
  & \$37.34 & \$3.05 \\
check
  & \$3.77  & \$0.32
  & \$13.39 & \$1.08
  & \$2.93  & \$0.25
  & \$5.44  & \$0.45
  & \$7.86  & \$0.64 \\
libfixmath
  & \$1.17  & \$0.11
  & \$2.95  & \$0.24
  & \$1.72  & \$0.15
  & \$1.29  & \$0.11
  & \$7.31  & \$0.60 \\
\bottomrule
\end{tabular}%
}
\end{table}

\paragraph{Analysis.}
Phase~1 (C documentation) dominates cost for large codebases such as \emph{stb}
(\$21.17/run on GPT-5.4), which contains over 1.9M prompt tokens owing to its
single-file, all-in-one header library design.
Phase~2 (translation) shows the highest variance: \emph{klib} requires
\$87.80 on GPT-5.4 due to its template-heavy macro system requiring many
planning-and-fix iterations, while simpler repositories such as \emph{libplist}
and \emph{libfixmath} cost under \$3.
Phase~5 (execution-aware refinement) is consistently expensive per
iteration---roughly \$7--\$37 on GPT-5.4---because each iteration involves
full repository context and long backtrace output from failing tests.
Across all stages Kimi-K2-Instruct costs approximately 7--9\% of the
equivalent GPT-5.4 run, offering a substantial price-performance trade-off
for large-scale migration workloads.

%% file: sections/appendix/EF_limitations_broader.tex
\newpage
\section{Broader Impacts}
\label{sec:appendix-broader}






\subsection{Security benefits.}
Memory-safety vulnerabilities in C codebases (buffer overflows, use-after-free,
data races) are responsible for a large fraction of reported CVEs.
Automated C-to-Rust migration directly reduces the attack surface of legacy
software by replacing unsafe memory management with Rust's ownership model.
\tool{}'s \code{unsafe}-minimisation loop further ensures that the resulting Rust
code retains idiomatic safety guarantees rather than mechanically wrapping C
patterns in \code{unsafe} blocks.

\subsection{Democratisation of code migration.}
Manual C-to-Rust porting requires deep expertise in both languages and is
prohibitively expensive for many organisations.
\tool{} lowers this barrier by automating the most labour-intensive steps
(documentation extraction, skeleton generation, iterative refinement), enabling
smaller teams and open-source projects to benefit from Rust's safety properties
without dedicated migration engineers.

\subsection{Open-source release.}
We plan to release \tool{}, CodeWikiBench, and all evaluation scripts under the
MIT licence.
This includes the full agent prompts, evaluation rubrics, and translated
repository snapshots used in our experiments, facilitating reproducibility and
enabling the community to extend the benchmark to additional repositories.

\subsection{Ethical considerations.}
Automated code migration carries the risk of subtle functional regressions that
pass compilation and even unit tests yet silently alter program behaviour.
Practitioners should treat \tool{}'s output as a first-pass migration aid subject
to human code review rather than a fully autonomous replacement for expert
engineering.
Furthermore, LLM-generated code may inadvertently reproduce copyrighted patterns
from training data; users should review the generated Rust code with appropriate
legal diligence before redistribution.

%% file: checklist.tex
\section*{NeurIPS Paper Checklist}

\begin{enumerate}

\item {\bf Claims}
    \item[] Question: Do the main claims made in the abstract and introduction accurately reflect the paper's contributions and scope?
    \item[] Answer: \answerYes{}
    \item[] Justification: The abstract and Section~\ref{sec:introduction} state the four contributions claimed by the paper (a documentation-driven migration paradigm, a documentation-guided iterative refinement mechanism, an execution-aware code revision stage with test-driven feedback, and an evaluation on eight large-scale real-world C repositories), each of which is realised by Section~\ref{doc-guided}--\ref{execution-aware} and supported by the empirical results in Section~\ref{sec:eval-compile}--\ref{sec:eval-safety}.

\item {\bf Limitations}
    \item[] Question: Does the paper discuss the limitations of the work performed by the authors?
    \item[] Answer: \answerYes{}
    \item[] Justification: The limitations of this work are discussed in Section~\ref{sec:limitation}.

\item {\bf Theory assumptions and proofs}
    \item[] Question: For each theoretical result, does the paper provide the full set of assumptions and a complete (and correct) proof?
    \item[] Answer: \answerNA{}
    \item[] Justification: The paper presents an empirical, agentic framework and does not contain formal theoretical results. The single equation in Section~\ref{requirement-refine} ($\texttt{CodeEquiv}(S,T) \approx \texttt{DocEquiv}(S_{\mathrm{doc}}, T_{\mathrm{doc}})$) is an explicit modelling approximation rather than a theorem.

\item {\bf Experimental result reproducibility}
    \item[] Question: Does the paper fully disclose all the information needed to reproduce the main experimental results of the paper to the extent that it affects the main claims and/or conclusions of the paper (regardless of whether the code and data are provided or not)?
    \item[] Answer: \answerYes{}
    \item[] Justification: Section~\ref{doc-guided}--\ref{execution-aware} fully specifies the agent roles (\emph{Planner}, \emph{Translator}, \emph{Synthesizer}, \emph{RequirementRefiner}, \emph{TestTranslator}, \emph{ExecutionRevisor}) and the tools each agent uses; Section~\ref{sec:eval-compile}--\ref{sec:eval-safety} specifies the dataset, metrics (Project Compilability, FCV, TPR, SR), baselines, and the two model backbones (Kimi-K2-Instruct, GPT-5.4). We additionally plan to release the framework code and the curated benchmark.

\item {\bf Open access to data and code}
    \item[] Question: Does the paper provide open access to the data and code, with sufficient instructions to faithfully reproduce the main experimental results, as described in supplemental material?
    \item[] Answer: \answerYes{}
    \item[] Justification: We plan to release both the \tool framework implementation and the curated eight-repository C benchmark, together with scripts to reproduce the reported metrics. An anonymised URL will be provided as supplemental material during the review period, and the assets will be made publicly available upon acceptance.

\item {\bf Experimental setting/details}
    \item[] Question: Does the paper specify all the training and test details (e.g., data splits, hyperparameters, how they were chosen, type of optimizer) necessary to understand the results?
    \item[] Answer: \answerYes{}
    \item[] Justification: Section~\ref{sec:eval-compile}--\ref{sec:eval-safety} reports the dataset (eight C repositories with maintainer test suites), evaluation protocol (cross-evaluated test suites \textbf{R} and \textbf{C}), metrics (Project Compilability, FCV, TPR, SR), baselines (C2Rust, Self-Repair, EvoC2Rust, Claude Code), and the two model backbones used. Our setting does not involve gradient-based training, so we do not report optimizer hyperparameters.

\item {\bf Experiment statistical significance}
    \item[] Question: Does the paper report error bars suitably and correctly defined or other appropriate information about the statistical significance of the experiments?
    \item[] Answer: \answerNo{}
    \item[] Justification: Each (method $\times$ model $\times$ repository) configuration is evaluated with a single run, so we report point estimates without error bars. Producing meaningful error bars would require repeating the entire multi-agent migration pipeline across multiple seeds, which is prohibitive given the LLM-inference cost of repository-scale translation reported in Appendix~\ref{sec:appendix-cost}.

\item {\bf Experiments compute resources}
    \item[] Question: For each experiment, does the paper provide sufficient information on the computer resources (type of compute workers, memory, time of execution) needed to reproduce the experiments?
    \item[] Answer: \answerYes{}
    \item[] Justification: Appendix~\ref{sec:appendix-cost} reports a per-stage LLM cost breakdown (USD) for each of the eight benchmark repositories under both Kimi-K2-Instruct and GPT-5.4, covering all five pipeline stages (C-Doc, Translation, Rust-Doc, Doc-Refine, Exec-Refine), with the pricing assumptions used for the estimate stated in the same section. The dominant cost of the framework is LLM API inference rather than local compute; \texttt{cargo} build/test executions run on a standard developer workstation.

\item {\bf Code of ethics}
    \item[] Question: Does the research conducted in the paper conform, in every respect, with the NeurIPS Code of Ethics \url{https://neurips.cc/public/EthicsGuidelines}?
    \item[] Answer: \answerYes{}
    \item[] Justification: We have reviewed the NeurIPS Code of Ethics. The work uses publicly available open-source C repositories and standard LLM APIs, does not involve human subjects, and preserves anonymity in this submission.

\item {\bf Broader impacts}
    \item[] Question: Does the paper discuss both potential positive societal impacts and negative societal impacts of the work performed?
    \item[] Answer: \answerYes{}
    \item[] Justification: Broader impacts are discussed in Appendix~\ref{sec:appendix-broader}. Positive impacts include reducing memory-safety vulnerabilities in legacy C infrastructure by automating migration to Rust, and democratising C-to-Rust porting for smaller teams and open-source projects. We also discuss ethical considerations: automated migrations may introduce subtle behavioural regressions that pass compilation and unit tests, so \tool's output should be treated as a first-pass migration aid subject to human code review, and LLM-generated code may inadvertently reproduce copyrighted patterns and warrants legal diligence before redistribution.

\item {\bf Safeguards}
    \item[] Question: Does the paper describe safeguards that have been put in place for responsible release of data or models that have a high risk for misuse (e.g., pre-trained language models, image generators, or scraped datasets)?
    \item[] Answer: \answerNA{}
    \item[] Justification: The paper does not release a pre-trained generative model, an image generator, or a scraped dataset. The released artefacts are a translation framework and a benchmark composed of already-public open-source C repositories.

\item {\bf Licenses for existing assets}
    \item[] Question: Are the creators or original owners of assets (e.g., code, data, models), used in the paper, properly credited and are the license and terms of use explicitly mentioned and properly respected?
    \item[] Answer: \answerYes{}
    \item[] Justification: All existing assets are properly cited: C2Rust~\cite{c2rust}, Self-Repair~\cite{c2rust-bench}, EvoC2Rust~\cite{Wang2025EVOC2RUSTASI}, CodeWiki~\cite{hoang2025codewiki}, Claude Code~\cite{santos2025claudecode}, Kimi-K2-Instruct~\cite{kimi2025k2}, and GPT-5.4~\cite{openai2025gpt5}, together with the eight public C repositories used for evaluation. All baselines and repositories are open-source projects used under their respective permissive licenses, and the LLM APIs are used in accordance with their providers' terms of service.

\item {\bf New assets}
    \item[] Question: Are new assets introduced in the paper well documented and is the documentation provided alongside the assets?
    \item[] Answer: \answerYes{}
    \item[] Justification: We plan to release the \tool framework and the curated eight-repository C benchmark, accompanied by a README, license file, and instructions for reproducing the reported metrics. An anonymised URL will be included as supplemental material during the review period.

\item {\bf Crowdsourcing and research with human subjects}
    \item[] Question: For crowdsourcing experiments and research with human subjects, does the paper include the full text of instructions given to participants and screenshots, if applicable, as well as details about compensation (if any)?
    \item[] Answer: \answerNA{}
    \item[] Justification: The paper does not involve crowdsourcing or any research with human subjects.

\item {\bf Institutional review board (IRB) approvals or equivalent for research with human subjects}
    \item[] Question: Does the paper describe potential risks incurred by study participants, whether such risks were disclosed to the subjects, and whether Institutional Review Board (IRB) approvals (or an equivalent approval/review based on the requirements of your country or institution) were obtained?
    \item[] Answer: \answerNA{}
    \item[] Justification: The paper does not involve crowdsourcing or research with human subjects, so no IRB or equivalent review was required.

\item {\bf Declaration of LLM usage}
    \item[] Question: Does the paper describe the usage of LLMs if it is an important, original, or non-standard component of the core methods in this research? Note that if the LLM is used only for writing, editing, or formatting purposes and does \emph{not} impact the core methodology, scientific rigor, or originality of the research, declaration is not required.
    \item[] Answer: \answerYes{}
    \item[] Justification: LLMs are a core component of the proposed methodology: each agent in \tool (\emph{Planner}, \emph{Translator}, \emph{Synthesizer}, \emph{RequirementRefiner}, \emph{TestTranslator}, \emph{ExecutionRevisor}) is instantiated with an LLM backbone, and we explicitly evaluate two backbones (Kimi-K2-Instruct and GPT-5.4) as described in Section~\ref{doc-guided}--\ref{execution-aware} and Section~\ref{sec:eval-compile}--\ref{sec:eval-safety}.

\end{enumerate}